\documentclass[superscriptaddress, nobibnotes, reprint,floatfix, %linenumbers%,longbibliography,endfloats
]{revtex4-2}
\usepackage{upgreek}
\usepackage{amsmath,braket}
\usepackage{graphicx}
\usepackage{mathtools}
\usepackage{color}
\usepackage{xcolor}
\usepackage{setspace}
\usepackage{lineno}
\newcommand{\RNum}[1]{\uppercase\expandafter{\romannumeral #1\relax}}
\usepackage{comment}
\usepackage{soul}

\newcommand{\zy}{\textcolor{black}}
\newcommand{\xj}{\textcolor{black}}

\begin{document}

% \preprint{APS/123-QED}

\title{Coherence in resonance fluorescence}

\author{Xu-Jie~Wang}
\affiliation{Beijing Academy of Quantum Information Sciences, Beijing 100193, China}
\author{Guoqi~Huang}
\affiliation{Beijing Academy of Quantum Information Sciences, Beijing 100193, China}
\affiliation{School of Science, Beijing University of Posts and Telecommunications, Beijing 100876, China}
\author{Ming-Yang~Li}
\author{Yuan-Zhuo~Wang}
\affiliation{National Laboratory of Solid State Microstructures and School of Physics, Collaborative 
Innovation Center of Advanced Microstructures, Nanjing University, Nanjing 210093, China}
\author{Li~Liu}
\affiliation{Beijing Academy of Quantum Information Sciences, Beijing 100193, China}
\author{Bang~Wu}\thanks{E-mail:\\ wubang@baqis.ac.cn\\hlyin@ruc.edu.cn \\yuanzl@baqis.ac.cn}
% \email{wubang@baqis.ac.cn}
\affiliation{Beijing Academy of Quantum Information Sciences, Beijing 100193, China}
\author{Hanqing~Liu}
\author{Haiqiao~Ni}
\author{Zhichuan~Niu}
\affiliation{Key Laboratory of Optoelectronic Materials and Devices, Institute of Semiconductors, Chinese Academy of Sciences, Beijing 100083, China}
\affiliation{Center of Materials Science and Optoelectronics Engineering, University of Chinese Academy of Sciences, Beijing 100049, China}
\author{Weijie~Ji}
\affiliation{Beijing Academy of Quantum Information Sciences, Beijing 100193, China}
\author{Rongzhen Jiao}
\affiliation{School of Science, Beijing University of Posts and Telecommunications, Beijing 100876, China}
\author{Hua-Lei~Yin}\thanks{E-mail:\\ wubang@baqis.ac.cn\\hlyin@ruc.edu.cn \\yuanzl@baqis.ac.cn}
% \email{hlyin@ruc.edu.cn}
\affiliation{Beijing Academy of Quantum Information Sciences, Beijing 100193, China}
\affiliation{National Laboratory of Solid State Microstructures and School of Physics, Collaborative 
Innovation Center of Advanced Microstructures, Nanjing University, Nanjing 210093, China}
\affiliation{School of Physics and Beijing Key Laboratory of Opto-electronic Functional Materials and Micro-nano Devices, Key Laboratory of Quantum State Construction and Manipulation (Ministry of Education), Renmin University of China, Beijing 100872, China}
\author{Zhiliang~Yuan}\thanks{E-mail:\\ wubang@baqis.ac.cn\\hlyin@ruc.edu.cn \\yuanzl@baqis.ac.cn}
% \email{yuanzl@baqis.ac.cn}
\affiliation{Beijing Academy of Quantum Information Sciences, Beijing 100193, China}
\date{\today}% It is always \today, today,

\begin{abstract}%Number of words:141<150
\textbf{Resonance fluorescence of a two-level emitter displays persistently anti-bunching irrespective of the excitation intensity, but inherits the driving laser’s linewidth under weak monochromatic excitation. These properties are commonly explained in terms of two \textit{disjoined} pictures, i.e., the emitter’s single photon saturation or passively scattering light. Here, we propose a unified model that treats all fluorescence photons as spontaneous emission, one at a time, and can explain simultaneously both the spectral and correlation properties of the emission.  
We theoretically derive the excitation power dependencies, measurable at the single-photon incidence level, of the first-order coherence of the whole resonance fluorescence and super-bunching of the spectrally filtered, followed by experimental confirmation on a semiconductor quantum dot micro-pillar device. Furthermore, our model explains peculiar coincidence bunching observed in phase-dependent two-photon interference experiments.  Our work provides an \zy{intuitive} understanding of coherent light-matter interaction and may stimulate new applications.}

%Keywords: Entanglement, coherence, superposition multilayered, quantum superposition, interference 

\end{abstract}
\maketitle

%A two-level emitter (TLE) under weak monochromatic excitation scatters out photons that are anti-bunched and yet exhibit the driving laser’s linewidth~\cite{hoffges1997heterodyne}. While this textbook experiment has been be accounted for by the formalism of light-atom interaction~\cite{scully_zubairy_1997,mandel1995optical,loudon2000quantum, Steck2023}, this simplest form of resonance fluorescence (RF) continues to fanscinate~\cite{Lopez_Carreno_2018,phillips2020,Hanschke2020,masters2023,casalengua2024two, Ng_2022,wu2023}  due to its potential applications~\cite{nguyen2011ultra,Konthasinghe2012, mattiesen2012,matthiesen2013phase} in generation of indistinguishable single photons as well as its intriguing physical explanations~\cite{Lopez_Carreno_2018,phillips2020,Hanschke2020,masters2023,casalengua2024two, Ng_2022,wu2023}. 

Although being a textbook phenomenon~\cite{scully_zubairy_1997,mandel1995optical, loudon2000quantum, Steck2023},
%fox_quantum_2006,
%grynberg_introduction_2010
resonance fluorescence (RF) continues to be an active research topic even in its simplest form~\cite{Lopez_Carreno_2018,phillips2020,Hanschke2020,masters2023,casalengua2024two, Ng_2022,wu2023}, where a two-level emitter (TLE) under weak monochromatic excitation scatters out photons that are anti-bunched and yet exhibit the driving laser’s linewidth~\cite{hoffges1997heterodyne, nguyen2011ultra,Konthasinghe2012, mattiesen2012,matthiesen2013phase}. While both phenomena can be calculated by the unwavering formalism of light-atom interaction~\cite{scully_zubairy_1997,mandel1995optical,loudon2000quantum,Steck2023}, confusion remains in their interpretations.  
Usually, anti-bunching is interpreted in the single-photon picture~\cite{Carmichael_1976, kimble1977}, where a TLE absorbs and re-emits photons one at a time. However, this picture fails to account for the RF's linewidth, which is far narrower than the natural broadening ($\frac{1}{2\pi T_1}$) imposed by the emitter's radiative lifetime ($T_1$). Conversely, it is easy to explain the laser-like spectrum if one treats the TLE only as a passive scattering site~\cite{lodahl2015}, but explaining consistently the single-photon characteristic becomes challenging. %Even more perplexing is the fact that the RF spectrum changes dramatically when the excitation intensity increases over a large range, but the single-photon characteristic persists~\cite{Wrigge2008, flagg_2009,ates2009,Ng_2022,wu2023}, leading to greater difficulties with the two disjoined scattering pictures.

Accompanying the spectrally sharp peak, RF spectrum of a TLE contains also a broadband component, which is vanishingly insignificant under weak excitation (Heitler regime) but grows towards dominance and eventually develops into Mollow triplets under strong excitation~\cite{mollow1969}.
Recently, L\'opez Carre\~no et al.~\cite{Lopez_Carreno_2018} theoretically clarified that the broadband component, however insignificant it may be, that holds the key to the presence or disappearance of anti-bunching through interference with the laser-like spectral component. 
Drastically departing from the single-photon picture, the broadband component was attributed to higher-order scattering processes involving ``the actual \textit{two-photon} absorption and re-emission"~\cite{Lopez_Carreno_2018,casalengua2024two}, which prompted experiments on the spectral filtering's effect on the photon number statistics~\cite{phillips2020,Hanschke2020, masters2023} and even led to suggestion of simultaneous scattering of two photons by an atom~\cite{masters2023}.   
However, spectrally resolving the RF  
invokes interference among (many) photons emitted over a macroscopic duration no less than $1/\Delta \nu$ ($\Delta \nu$: spectral resolution), which would prevent discerning what truly happens at the minuscule duration of the TLE's radiative lifetime.  
From the perspective of wave-particle duality, the action of filtering reveals already the wave-aspect properties, and thus undermines discussions on simultaneity which requires treating photons as particles.

\begin{figure*}[tb]		
\centering
\includegraphics[width=1.6\columnwidth]{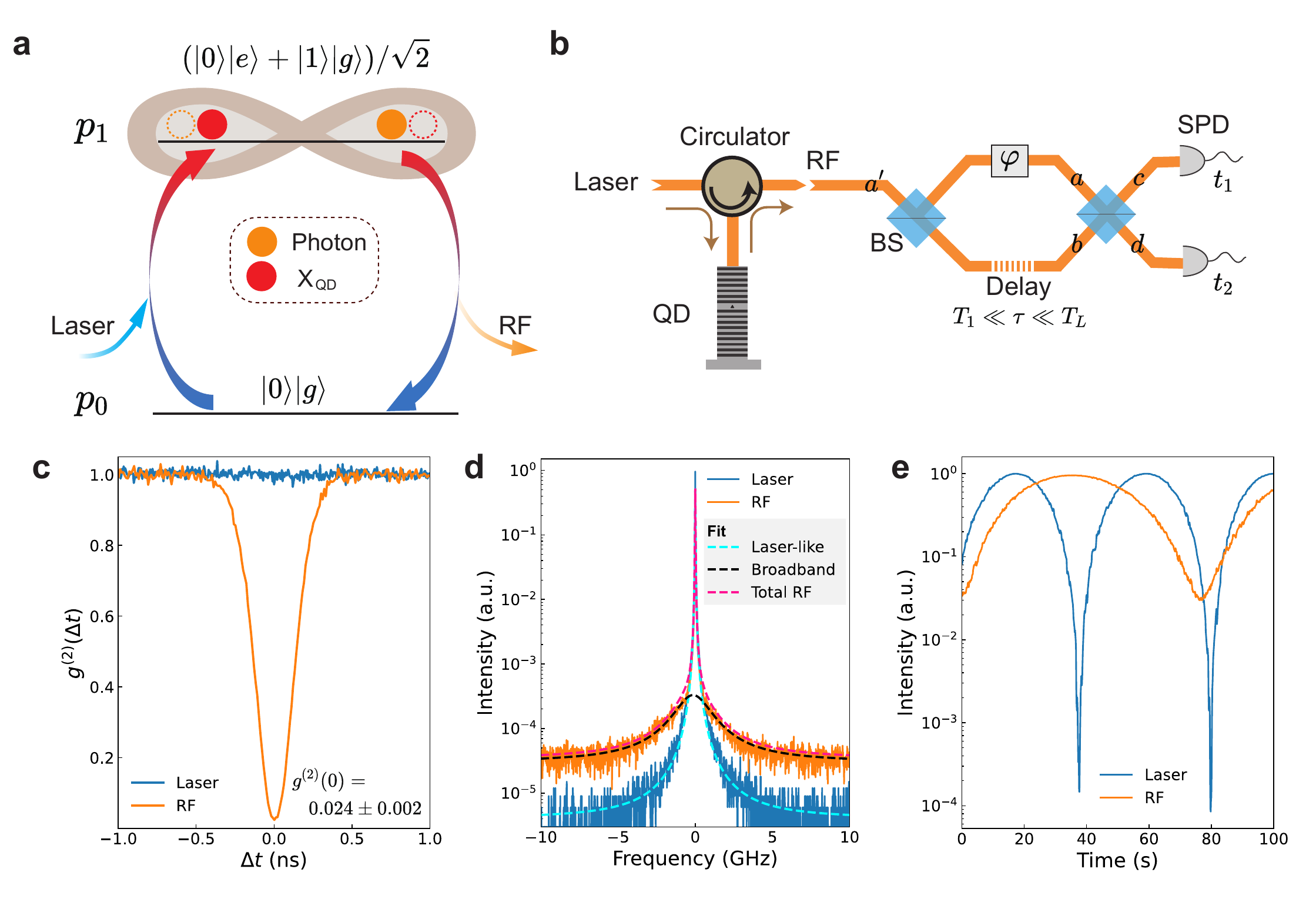}	
\caption{\textbf{Resonance fluorescence (RF).} \textbf{a}, Schematic for a two-level emitter (TLE) coherently driven by a continuous-wave laser into steady-state. Brackets
$\ket{g}$ and $\ket{e}$ represent the ground and excited states of the TLE, while $\ket{0}$ and $\ket{1}$ mean 0 or 1 spontaneously emitted photon.  Symbols $p_0$ represents the population of the system ground while $p_1$ is the single-quanta population that is in a form of either the TLE staying at its excited state ($\ket{0}\ket{e})$) or a fresh spontaneously emitted photon ($\ket{1}\ket{g}$).  
 \textbf{b}, Schematic of the core experimental setup. The RF is collected in the same polarization as the excitation laser.
  \textbf{c}, Second-order correlation function $g^{(2)}(\Delta t)$ measurements.  
 \textbf{d}, High resolution spectra. \textbf{e},  Interference fringes measured with the AMZI shown in \textbf{b}. In this measurement we left the AMZI's  phase $\varphi$ drift freely and recorded the count rate of the SPDs as a function of elapsed time. BS, beam splitter, SPD, single-photon detector.
 }
	\label{fig1}
\end{figure*}

In this work, we propose and experimentally verify  a unified model 
which simultaneously explains the laser-like linewidth and the single-photon characteristics of the RF of a quantum emitter. 
The model involves no higher-order scattering processes, nor does it need to distinguish between ``coherent" scattering and ``incoherent" absorption/re-emission processes. 
\textcolor{black}{Here, all RF photons are strictly treated as spontaneous emission, one at a time and never two together.}
We theoretically derive the excitation power dependencies, with the strongest effects measurable at the single-photon incidence level, of the first-order coherence ($g^{(1)}$) of the RF as whole and the second-order correlation function ($g^{(2)}$) of its broadband component. In laboratory, we confirm the model by reproducing its predictions on the RF from a high-quality semiconductor quantum dot (QD) micro-pillar device.

\noindent\textbf{\large Results}\\ 
Figure~\ref{fig1}\textbf{a} depicts the \zy{spontaneous emission} model. A monochromatic laser coherently drives a TLE, e.g., a QD in a micropillar cavity~\cite{wu2023}. Parameters $\nu$, $h \nu$ and $T_L$ are the laser's frequency, photon energy and coherence time, respectively.  
The TLE's excited state $|e\rangle$ has a lifetime of $T_1$, and a dephasing time of $T_2$ ($T_2 \leq 2T_1$).  
According to the optical Bloch equations, the TLE population reaches a steady state after $\sim\! T_1$ time under steady continuous optical excitation~\cite{Steck2023}.
\zy{Fundamentally, the TLE or its emission separately is in a mixed state.  However, treating them jointly, it is possible, as derived in Section~I, Supplementary Information, to use a pure-state description} 
\begin{equation}
       |\psi\rangle_t = \sqrt{p_0}|0\rangle_t|g\rangle_t + \sqrt{p_1} e ^ {\zy{-\textbf{i}2\pi \nu t}} \frac{\ket{0}_t\ket{e}_t + \ket{1}_t \ket{g}_t}{\sqrt{2}},
\label{eq:composite}
\end{equation}
\noindent where $\sqrt{p_0}$ and  $\sqrt{p_1}$ represent the \zy{magnitudes of the quantum probability amplitudes} ($p_0 + p_1 = 1$). 
%$\ket{1}_t$ means the spontaneously emitted photon. 
Here, $|0\rangle_t|e\rangle_t$ means the \zy{TLE} occupying its excited state has not spontaneously emitted at  time $t$, while $|1\rangle_t|g\rangle_t$ indicates emission of a photon has just taken place with the \zy{TLE} having returned to its ground state $|g\rangle_t$.  
%We clarify that $\ket{0}_t$ and $\ket{1}_t$ here shall not be mistaken as cavity photons as in a Jaynes-Cummings-like model~\cite{Steck2023,fox_quantum_2006,grynberg_introduction_2010}.
States $\ket{0}_t\ket{e}_t$ and $\ket{1}_t\ket{g}_t$ are connected only via spontaneous emission, and $\ket{1}_t$ represents a spontaneously emitted photon at  $t$ contributing to the RF. By definition, $\ket{1}_t$ is a broadband photon with bandwidth governed by the TLE's dephasing time $T_2$. 
\zy{The state $\frac{\ket{0}_t\ket{e}_t + \ket{1}_t \ket{g}_t}{\sqrt{2}}$ implies that, }
at any  given instant of time,  it is not possible to know whether spontaneous emission has taken place or the TLE remains at its excited state $\ket{e}$ prior to detection of a photon.  However, once an RF photon is detected, we know for sure that spontaneous emission had taken place and the TLE returned to its ground state $|g\rangle$ at the corresponding \zy{instant} of time.  An immediate detection of a second photon is prevented because the TLE requires time to be repopulated. This leads naturally to photon anti-bunching. 

%
%Below, we show how the entanglement can transfer partially its coherence to RF using \textsl{two-path} interference. 
%
%\zy{Compared to spectral filtering~\cite{phillips2020,Hanschke2020,masters2023} that is multi-path interference, the simplicity of two-path interference allows to apply Eq.~\ref{eq:composite} in the subsequent analysis.} 
Figure~\ref{fig1}\textbf{b} shows an asymmetrical Mach-Zehnder interferometer (AMZI) suitable to evaluate the RF's coherence.
The AMZI's delay ($\tau$) is chosen to be much longer than the TLE relaxation time, but much shorter than the coherence time of the excitation laser, i.e., $T_1 \ll \tau \ll T_L$, so as to ensure the state-state, \zy{and phase-coherent,} condition.
The incoming RF signal from port $a^\prime$ is divided equally into two paths ($a$ and $b$) by the first beam splitter and then recombine at the second one before detection at ports $c$ and $d$ by two single-photon detectors. 
When a photon is detected, it is not possible to distinguish whether it arose from the one emitted to an early time ($t-\tau$) taking the long path or one emitted at a late time ($t$) taking the short path.  Interference between the two indistinguishable paths occurs.
\zy{As the coherence properties of an RF signal is unaffected by channel losses  (see Section II, Supplementary Information),  we can then  use the pure-state of Eq.~\eqref{eq:composite} as the input to  analyse the interference outcome after the AMZI. The AMZI output state can be written as} 
\begin{widetext}
{\begin{equation}
\begin{aligned}%\label{intfer}
\ket{\Psi_{\rm out} }= &\ket{0_c0_d}_t\left(p_{0}\ket{gg}+\frac{\sqrt{p_{0}p_{1}}}{\sqrt{2}}\ket{ge}+\frac{\sqrt{p_{0}p_{1}}}{\sqrt{2}}\ket{eg}+\frac{p_{1}}{2}\ket{ee}\right)\\
+&\ket{1_c0_d}_t\frac{\sqrt{p_{0}p_{1}}}{\sqrt{2}}\frac{1-e^{\textbf{i}\varphi}}{\sqrt{2}}\ket{gg}
+\ket{1_c0_d}_{t}\frac{p_{1}}{2\sqrt{2}}\left(\ket{ge}-e^{\textbf{i}\varphi}\ket{eg} \right) -\ket{2_c0_d}_t e^{\textbf{i}\varphi}\frac{p_{1}}{2\sqrt{2}}\ket{gg}\\
+&\ket{0_c1_d}_t\frac{\sqrt{p_{0}p_{1}}}{\sqrt{2}}\frac{1+e^{\textbf{i}\varphi}}{\sqrt{2}}\ket{gg}+\ket{0_c1_d}_{t}\frac{p_{1}}{2\sqrt{2}}\left(\ket{ge}+e^{\textbf{i}\varphi}\ket{eg} \right)  +\ket{0_c2_d}_t e^{\textbf{i}\varphi}\frac{p_{1}}{2\sqrt{2}}\ket{gg},\\
\label{eq:output}
\end{aligned}
\end{equation}}
\end{widetext}
\noindent where $\varphi$ denotes the AMZI phase delay and $\ket{xy}$ ($x,y = g, e$) represents the TLE's respective states corresponding to time bins $t-\tau$ and $t$.  
The first line contains no photons while the second and third lines represent photon outputs at ports $c$ and $d$, respectively. 
Each output contains one phase-dependent term followed by two phase-independent ones. 
The phase-dependent term corresponds to the TLE's transition $(\ket{ge} + \ket{eg})/\sqrt{2} \to \ket{gg}$, which imparts the coherence to a superposition between two photon temporal modes: $(\ket{0}_{t-\tau}\ket{1}_{t}+\ket{1}_{t-\tau}\ket{0}_{t})/\sqrt{2}$. Varying the AMZI phase $\varphi$, this superposition will produce interference fringes with an amplitude of $\frac{p_0p_1}{2}$, as opposed to the total  output intensity of $\frac{p_1}{2}$. Thus, the coherence level of the RF, quantified using the first-order correlation function $g^{(1)}(\tau)$, has the form,
\begin{equation}
g^{(1)}(\tau) =  p_0 e^{-\textbf{i}2\pi \nu \tau}.
\label{eq:g1}
\end{equation}
Using Fourier transform, we infer that the RF consists of a spectrally sharp, laser-like ($ll$) part that inherits the linewidth of the driving laser and has a spectral weight of $I_{ll}/I_{tot} = |g^{(1)}(\tau)| =  p_0 < 1$.  For detailed theoretical derivation, see Sections II-IV, Supplementary Information.

\zy{The reduction in coherence, by the amount of $1-p_0$ or $p_1$, is linked to the TLE's transitions from the $\ket{ee}$ state to $\ket{ge}$, $\ket{eg}$ or $\ket{gg}$. 
As shown in Eq.~(\ref{eq:output}), the first two transitions give rise to an incoherent single photon each while the last one produces a two-photon state.
Transition $\ket{ee}\to\ket{ge}$ ($\ket{ee}\to\ket{eg}$) emitted a photon into early (late) temporal mode but none at late (early) mode, so no two-path interference takes place.  
On the other hand, transition $\ket{ee}\to\ket{gg}$ produces one photon into each mode, the interference of which causes coalescence and forms a photon-pair through Hong-Ou-Mandel (HOM) effect~\cite{Hong1987}.  
All these photons are incoherent, so they naturally display a bandwidth governed by the TLE's transition and thus make up  the broadband ($bb$) part, with a  weight of $I_{bb}/I_{tot} = p_1$ in the RF spectrum.}

In the absence of pure dephasing ($T_2 = 2T_1$), the relation of $p_0$ and $p_1$ with excitation power can readily be estimated through steady-state %\st{condition} 
equilibrium.  
We define $\bar{n}$ as the mean incident photon number over $T_1$ duration~\cite{wu2023}, and $\eta_{ab}$ as the TLE's absorption %\st{/re-emission (scattering)}
efficiency under weak excitation limit. 
With absorption balancing out emission, we obtain $\frac{\bar{n}}{T_1} \times \eta_{ab} \times (1 - p_1) = \frac{p_1}{2} \times  \frac{1}{T_1}$, where the factor $(1-p_1)$ on the left takes saturation into account while $\frac{p_1}{2}$ on the right reflects on average only half of the excited state is in the matter form. We then have 
\begin{equation}
\begin{split}
     p_0 &=\frac{1}{1+2\bar{n}\eta_{ab}}, \\
     p_1 &=\frac{2\bar{n}\eta_{ab}}{1+2\bar{n}\eta_{ab}}.
\end{split}  \label{eq:p1}
\end{equation}
These equations connect $p_0$ and $p_1$ directly to the single photon excitation level, and are applicable to both  Heitler and Mollow excitation regimes. %\zy{Under strong pump limit ($\bar{n} \to \infty$), $p_0 = 0$ ($p_1 = 1$) and the laser-like RF fraction disappears.}
Since no assumption is made on the emitter type, we believe Eq.~\ref{eq:p1}
holds, under the condition $T_2 = 2T_1$, for all quantum two-level emitters, including a trapped atom~\cite{masters2023, Ng_2022} or ion~\cite{hoffges1997heterodyne}, a molecule~\cite{Wrigge2008}, and a semiconductor QD~\cite{nguyen2011ultra, mattiesen2012, matthiesen2013phase, Konthasinghe2012, ates2009,flagg_2009}. 
The absorption efficiency $\eta_{ab}$ may vary drastically among emitters, but 
%\zy{However}, a transmission dip of 11.5~\% was reported for a molecule under a freely propagating resonant excitation in a single-pass encounter without a cavity~\cite{Wrigge2008}. 
embedding an emitter into a cavity can enhance the light-matter interaction and could bring $\eta_{ab}$ close to unity. 
For a high-quality QD-micropillar device~\cite{de_santis_2017,Proux2015,wu2023}, we expect %$p_1 \simeq \frac{2\bar{n}}{1+2\bar{n}}$ and 
$\textcolor{black}{|g^{(1)}(\tau)|} \zy{ = p_0} 
\simeq  \frac{1}{1+2\bar{n}}$, which presents a unique test-point to our model. 

\zy{Eq.~\eqref{eq:output} yields another experimentally verifiable prediction.
Looking at the third line of this equation, the laser-like component, though being dominant under weak excitation, can be completely eliminated through destructive interference by setting the AMZI phase to $\varphi = \pi$.  Containing just the non-interfering single-photon term ($\ket{0_c 1_d}_t\left ( \ket{ge} - \ket{eg} \right)$ and one photon-pair term ($\ket{0_c 2_d}_t \ket{gg}$), 
the photon state of port $d$ resembles closely a two-photon state attenuated by a beam-splitter. As derived in Section V, Supplementary Information, the port $d$ output at $\varphi = \pi$ will exhibit excitation-flux-dependent super-bunching} 
\begin{equation}
%\begin{aligned}
g^{(2)}(0) %&=\frac{\bra{\psi_{d}} a^{\dag}a^{\dag}aa\ket{\psi_{d}}_{\varphi = \pi}}{\bra{\psi_{d}}a^{\dag}a\ket{\psi_{d}}_{\varphi = \pi}^{2}} %=\frac{(p_{1}/2)^{2}}{(p_{1}^{2}/4+p_{1}^{2}/4)^{2}}
=\frac{1}{p_{1}^{2}}\\
=\left(1 + \frac{1}{2\bar{n}\eta_{ab}}\right)^2.
%g^{(2)}(0) =  1/p_1^2 = \left(1 + \frac{1}{2\bar{n}\eta_{ab}}\right)^2.
%\end{aligned}
\end{equation}
At low pump limit ($\bar{n} \to 0$), the $g^{(2)}(0)$ value can theoretically be infinitely large.  At the opposite limit ($\bar{n} \to \infty$), the RF loses all its first-order coherence according to Eq.~\eqref{eq:g1} and the super-bunching disappears, i.e., $g^{(2)}\zy{(0)} = 1$. 
\zy{We remark that the predicted super-bunching is not observable with  an incoherently pumped TLE~\cite{incoherent}. }

%
%    Port D output state by Hualei
%
%{\begin{equation}
%\begin{aligned}
%\ket{\psi_{d}}_{\varphi = \pi}&=\left(2p_{0}p_{1}+\frac{5p_{1}^{2}}{8}\right)\ket{0}\bra{0}+\frac{p_{1}^{2}}{4}\ket{1}\bra{1}+\left(p_{0}\ket{0}-\frac{p_{1}}{2\sqrt{2}}\ket{2}\right)\left(p_{0}\bra{0}-\frac{p_{1}}{2\sqrt{2}}\bra{2}\right),\\
%\ket{\Psi}&=\left(2p_{0}p_{1}+\frac{p_{1}^{2}}{2}\right)\ket{0}\bra{0}+\left(p_{0}\ket{0}-\frac{p_{1}}{\sqrt{2}}\ket{2}\right)\left(p_{0}\bra{0}-\frac{p_{1}}{\sqrt{2}}\bra{2}\right),\\
%\end{aligned}
%\end{equation}}

%\hly{where we partially trace out the atomic system and port $c$. The output state then is a mixture state with the vacuum state, the single-photon state, and a superposition of vacuum and two-photon states.} 

%Figure~\ref{fig1}\textbf{b} shows the core experimental setup.
To test, we use a QD-micropillar device featuring a quality factor of 9350 and a low cavity reflectivity of 0.015. It is kept in a closed-cycle cryostat and the QD's neutral exciton is temperature-tuned into the cavity resonance at 13.6~K, emitting at $911.54$~nm. 
The setup is schematically shown in Fig.~\ref{fig1}\textbf{b}.
We use a confocal microscope setup equipped with a tunable continuous-wave (CW) laser of 100~kHz linewidth as the excitation source and an optical circulator made of a polarising beam splitter and a quarter-wave plate for collecting the RF in %a co-polarisation configuration
the same polarization of the driving laser~\cite{wu2023,bennett2016semiconductor}.  
The QD is characterised to have a Purcell enhanced lifetime of $T_1=67.2$~ps ($F_p \simeq 10$), corresponding to a natural linewidth of $\zy{\gamma_\parallel}/2\pi = 2.37$~GHz, which is 15 times narrower than the cavity mode ($\kappa/2\pi = 35$~GHz).
The RF is fed into a custom-built AMZI with 
a fixed delay of $\tau = 4.92$~ns for interference before detection by single photon detectors.
With additional apparatuses, the whole setup allows characterisations of high-resolution spectroscopy,  auto-correlation function $g^{(2)}(\Delta t)$, the first-order correlation function $g^{(1)}(\tau)$, and two-photon interference. 
All excitation fluxes used were  strictly calibrated via the incident optical power ($P_{in}$)
upon the sample surface using the relation of $\bar{n} = P_{in}T_1/h\nu$.
\xj{Detailed description of the experimental setup can be found in Methods.}
%For detailed description of the experimental setup, see Section~\xj{VII}, Supplementary Information. 

In the first experiment, we use a weak excitation flux of $\bar{n} = 0.0068$, corresponding to a Rabi frequency of $\Omega/2\pi =210$~MHz ($\sim \!0.09\zy{\gamma_\parallel}$).
Figure \ref{fig1}\textbf{c} shows the auto-correlation function $g^{(2)}(\Delta t)$ for both the RF (orange line) and the laser (blue line).  While the laser exhibits a flat $g^{(2)}$ because of its Poissonian statistics, the RF is strongly anti-bunched,  with $g^{(2)}(0)=0.024\pm0.002$,  at the 0-delay over a time-scale of $\sim T_1$, confirming that the QD scatters one photon at a time.

Figure \ref{fig1}\textbf{d} shows the RF frequency spectrum (orange line) measured with a scanning Fabry-P\'{e}rot interferometer (FPI).
It is dominated by a sharp line that overlaps the laser spectrum (blue line) with a linewidth that is limited by the FPI resolution (20~MHz). 
The RF contains additionally a broadband pedestal whose amplitude is over 3 orders of magnitude weaker.  
The overall spectrum can be excellently fit with two Lorentzians of 20~MHz and 2.3~GHz linewidths, shown as cyan and black dashed lines, respectively. The bandwidth of the latter closely matches the TLE's natural linewidth of $\zy{\gamma_\parallel}/2\pi$.   
Following the discussion surrounding Eq.~\ref{eq:g1}, we attribute the sharp feature to the interference outcome of the RF signal passing through the FPI.   
The spectral weight of this laser-like peak can be 
measured using our AMZI (Fig.~\ref{fig1}\textbf{b}), %which has a suitable delay that 
whose delay meets the steady-state condition $T_1 \ll \tau \ll T_L$.
An example result is shown in Fig.~\ref{fig1}\textbf{e}, which gives a fringe visibility, or the laser-like fraction, of 0.94 for the RF. As comparison, the laser signal exhibits 0.9998 interference visibility. %indicating deterioration in the RF's coherence by its broadband component.

%\zy{Two-path interference not only brings substantial simplicity (and clarity) into analysis of experimental results, but also allows retrieval of a source's coherence or spectral property with a suitable delay range as commonly used in Fourier-transform spectroscopy~\cite{zwiller_2004}.} With a delay that is orders of magnitude away from either the emitter's radiative lifetime or the laser's coherence time,  the AMZI we build allows precise manipulation of the laser-like RF component and thus enables us to verify an entanglement model that we propose to treat the RF entirely as spontaneous emission by the emitter. 

\begin{figure}[tb]
	\centering	\includegraphics[width=1\columnwidth]{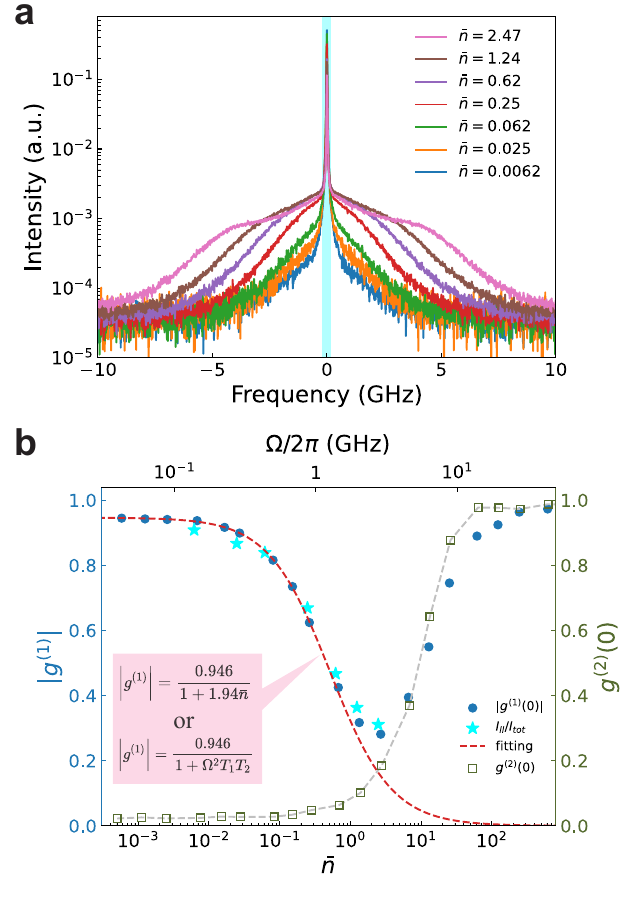}
	\caption{\textbf{Coherence versus incident flux.}  
 \textbf{a}, High-resolution spectra; The cyan bar illustrates the laser-like spectral part.
 \textbf{b},  $|g^{(1)}|$ (solid circles) measured with the AMZI and  $I_{ll}/I_{tot}$ (solid stars) extracted from data in panel \textbf{a}, as well as $g^{(2)}(0)$ measured without any spectral filtering.  
 The red dashed line is a fitting using either $|g^{(1)}| \propto \frac{1}{1+x\bar{n}}$ with $x = 1.94$ or $|g^{(1)}| \propto \frac{1}{1+ \Omega^2 T_1T_2}$ with $T_2 = 1. 62 T_1$. }
  	\label{fig2}
\end{figure}

Figure~\ref{fig2}\textbf{a} shows high-resolution RF spectra under different excitation fluxes. As the flux increases,  the broadband component increases its share of the total RF, and becomes considerably broadened when $\bar{n}$ exceeds 0.25. It starts to develop into Mollow triplets at the few photon level as reported previously~\cite{wu2023}. 
Nevertheless, the RF \xj{retains} its single-photon characteristics for a flux up to $\bar{n} = 6.8$ when measured before the AMZI and \textit{without any spectral filtering}, as demonstrated by the auto-correlation data (open squares) in Fig.~\ref{fig2}\textbf{b}. At $\bar{n} = 6.8$, we measure $g^{(2)}(0) = 0.37$, which is still below the limit (0.5) for a classical emitter. %The increase of $g^{(2)}(0)$ at high fluxes is caused by the laser background.

The growing broadband component deteriorates the RF's coherence. To quantify, we measure the interference visibility ($|g^{(1)}|$) by passing the RF through the AMZI, with results shown as solid circles in Fig.~\ref{fig2}\textbf{b}.  This quantification method is equivalent to, and more precise than, calculating the area ratio of the laser-like peak to the total RF signal. The results from the latter method are shown as stars. 
At low fluxes ($\bar{n} < 0.01$), $|g^{(1)}|$ is plateaued at 0.946, rather than 1.0 as expected from Eq.~\ref{eq:g1}. We attribute this discrepancy to photon distinguishability~\cite{loredo2019generation}, which could arise from phonon-scattering~\cite{iles-smith_2017, koong2019,brash2019} and QD environmental charge fluctuation~\cite{zhai2020} as well as a small amount of laser mixed into the RF. As the flux increases until $\bar{n} = 3$, we observe a general trend of a decreasing visibility.  For $\bar{n} > 3.0$, the visibility reverses its downward trend and climbs up.  In this regime, the RF signal starts to saturate~\cite{wu2023} while the laser background continues to rise, as evidenced by the accompanying rise in $g^{(2)}(0)$. At very strong fluxes ($\bar{n} > 100$), the laser background dominates because our setup  collects the RF in the same polarisation of the driving laser, and thus the measured photon number statistics approaches Poissonian distribution, i.e., $g^{(2)}(0) \approx 1$. 

\begin{figure*}[tb]
	\centering	\includegraphics[width=1.4\columnwidth]{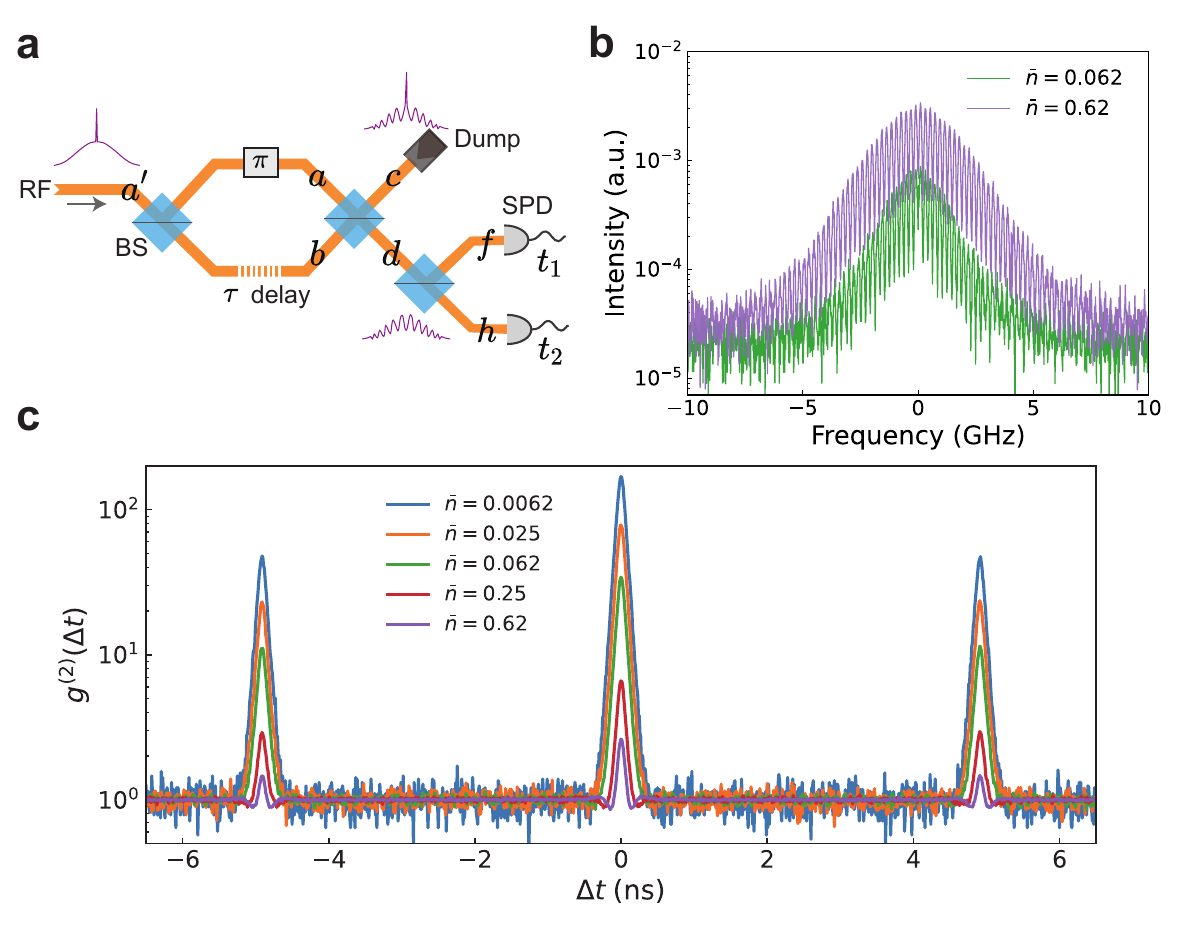}
	\caption{
 \textbf{Correlation of the AMZI-filtered RF.} \textbf{a}, Experimental setup; The AMZI is set to have a $\pi$ phase so that the laser-like RF component is dumped at port $c$; The filtered RF at port $d$ contains only the broadband component, and is fed into a HBT setup. \textbf{b}, Filtered RF spectra; \textbf{c}, Second-order correlation functions measured for various excitation fluxes.}
 	\label{fig3}
\end{figure*}

We attribute the interference visibility drop in Fig.~\ref{fig2}\textbf{b} to the increasing population ($p_1$) of the QD's exited state.  
Based on Eqs.~\ref{eq:g1} and \ref{eq:p1}, we obtain a near perfect fit, $|g^{(1)}| = 0.946 /(1+x\bar{n})$ and $x = 1.94$,  to the experimental data in the region where the RF signal remains dominant.
The fitting is identical to the result (also shown as dashed line) by the traditional model~\cite{loudon2000quantum,nguyen2011ultra,Proux2015}: $|g^{(1)}| \propto \frac{1}{1+\Omega^2 T_1T_2}$, where we use $T_2 = 1.62T_1$ and the Rabi frequencies extrapolated from the Mollow splittings measured under strong excitations. This is unsurprising because both models share the same theoretical foundation and $\bar{n} \propto \Omega^2$ holds.
However, our model gives a more direct link of the RF's coherence to the excitation power, without the need for extracting 
the Rabi frequencies in order to achieve so. 
This experiment strongly supports our model and at the same time suggests an efficient input-coupling of our QD-micropillar device with $\eta_{ab} \simeq 0.97$.

%%%%%%%%%%%%%%
In the next experiment (Fig.~\ref{fig3}\textbf{a}), we use the AMZI to filter out the laser-like component by setting its phase to $\varphi = \pi$ and then examine the super-bunching as expected in Eq.~4. %the photon number statistics of the filtered RF output. 
As compared with a narrow-band filter~\cite{phillips2020,Hanschke2020,masters2023}, this technique uses two-path, instead of multi-path, interference and thus the subsequent photon number statistics is easier to analyse. For the theoretical analysis, see Section V of Supplementary Information.
Two AMZI-filtered spectra are shown in Fig.~\ref{fig3}\textbf{b}.  Each spectrum consists of a broadband signal with interference fringes of 203~MHz spacing  corresponding to the AMZI's delay ($\tau = 4.92$~ns), while the laser-like component is rejected entirely.  
Subjecting the filtered RF to the auto-correlation measurement, we acquire a set of data shown in Fig.~\ref{fig3}\textbf{c}.  We observe super-bunching at 0-delay with $g^{(2)}(0) = 168.9$ at the lowest flux of $\bar{n} = 0.0062$.
At $\Delta t = \pm \tau$, interference between three temporal modes happens. Theoretically, $g^{(2)}(\pm \tau) \approx \frac{1}{4}g^{(2)}(0)$ under such incident flux. To compare, we have measured an average value of 47.6 for $g^{(2)}(\pm \tau)$, amounting to 0.282 of the corresponding $g^{(2)}(0)$ value.   

We attribute the observed super-bunching to a combined effect of two-photon interference and the RF's first-order destructive interference. The former generates two-photon states contributing to the 0-delay coincidences, while the latter reduces the photon intensity in the AMZI output port and thus suppresses the coincidence baseline.  
Setting $\varphi = \pi$ maximises the level of super-bunching, but $g^{(2)}(0)$ can be tuned continuously down to anti-bunching through the AMZI phase. \xj{Details about AMZI phase stabilization and }measurements for other characteristic phases are shown in \xj{Methods}.
%Section VIII, Supplementary Information.

As the excitation flux increases, we expect the RF's first-order coherence to reduce and so will the level of super-bunching. 
Figure~\ref{fig3}\textbf{c} shows the excitation flux dependence of the auto-correlation, which is in qualitative agreement with the theoretical prediction of Eq.~4.
At $\bar{n}=0.62$, we deduce $p_1= 0.546$ using the empirical relation of $p_1 = 1.94 \bar{n}/(1+1.94\bar{n})$ and thus expect a photon bunching value of 3.35. Experimentally, we obtain $g^{(2)}(0) = 2.6$, which is in fair agreement with the expected value. The discrepancy could arise from the increased laser background as well as finite photon indistinguishability~\cite{loredo2019generation}.  
%%%%%%%%%%%%%%

\begin{figure*}[t]
	\centering	\includegraphics[width=0.75\textwidth]{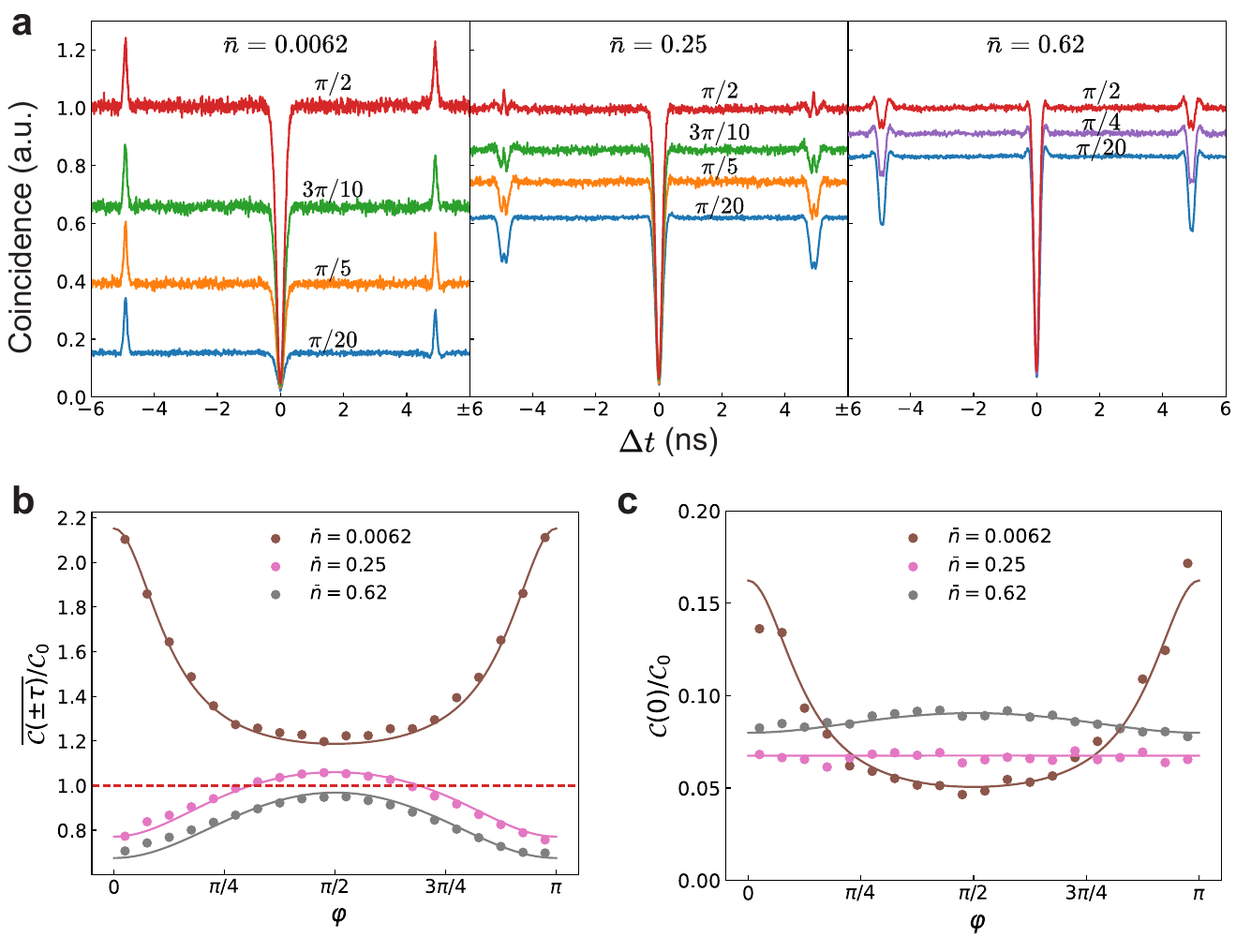}
	\caption{
 \textbf{Phase-dependent two-photon interference.}   
  \textbf{a}, Cross-correlation traces measured with  the setup shown in Fig.~\ref{fig1}\textbf{b} for three different excitation intensities: $\bar{n} = 0.0062$ (left), $0.25$ (middle) and $0.62$ (right panel).  \textbf{b}, 
  Measured (solid symbols) and theoretical (solid lines) coincidence probabilities at $\Delta t = \pm \tau$ delays, normalised to the coincidence baseline (dashed line).  We use $\overline{\mathcal{C}(\pm \tau)} = \frac{1}{2}\left( \mathcal{C}(+\tau) + \mathcal{C}(-\tau) \right)$.  \textbf{c}, Normalised experimental (solid symbols) and theoretical (solid lines) coincidence probabilities at  $\Delta t = 0$.  Experimental data 
  in panels \textbf{b} and \textbf{c} are extracted from data in panel \textbf{a}. The theoretical results are fitted using maximum likelihood estimation. Fitted  parameters $\{p_0, p_1,p_2; M^\prime\}$ corresponding to different excitation fluxes $\bar{n}=0.0062, 0.25, 0.62$ are $\{ 0.98,0.023,8.0 \times 10^{-6}; 0.96\}$, $\{0.69, 0.30, 2.2 \times 10^{-3}; 0.94\}$ and $\{0.49, 0.50, 8.0 \times 10^{-3}; 0.92\}$, respectively. 
  A fixed value of $M=0.89$, extracted from the plateaued $|g^{(1)}| = 0.946$ shown in Fig.~2\textbf{b} through $M = |g^{(1)}|^2$, is used 
  for photon indistinguishability.}
	\label{fig4}
\end{figure*}

Finally, we perform phase-dependent two-photon interference experiment with the setup shown in Fig.~\ref{fig1}\textbf{b}, and summarize the results in Fig.~\ref{fig4}\textbf{a} with observations: (1) The coincidence baseline is phase-dependent, while the gap between traces shrinks as the excitation power increases; (2) Strong anti-bunching at $\Delta t = 0$ for all excitation fluxes and phase values; (3) Features at $\Delta t = \pm 4.92$~ns, caused by the AMZI's delay $\tau$, can exhibit as peaks or dips depending on both the excitation power and the phase delay. 
We note that observation (3) is strikingly different from incoherently excited quantum emitters~\cite{Proux2015, PhysRevLett.100.207405}, where the side features always display as dips with depth limited to 0.75. 

To understand the two-photon interference results,  we approximate the RF output as a superposition of photon-number states:  $\ket{\psi_{ph}}_t = \sqrt{p_0}|0\rangle_t + \sqrt{p_1}|1\rangle_t + \sqrt{p_2}|2\rangle_t$ with a small two-photon probability $p_2 \ll p_1^2/2$ 
%($p_0+p_1+p_2= 1$) 
and derive the coincidence probabilities as detailed in Section VI of Supplementary Information.  We reproduce the main results below.  
%\begin{widetext}
\begin{equation}
\mathcal{C}(0)=\frac{p_2}{4}\left(1-p_0 M \cos2\varphi \right)+\frac{p_1^2+4p_1p_2+4p_2^2}{8}(1-M^\prime),  \tag{6A}
\end{equation}
\begin{equation}
\mathcal{C}(\pm\tau)= \frac{p_1^2}{16} (3 - 2 p_0%^{\textcolor{blue}{2}}
M\cos 2\varphi),  \tag{6B}
\end{equation}
\begin{equation}
\mathcal{C}_0 =\frac{p_1^2}{4}\left(1-p_0^2 M\cos^2\varphi\right),  \tag{6C}
\end{equation}
%\end{widetext} 
\noindent where $M$ represents indistinguishability of the RF photons while $M^\prime$ is the post-selective two-photon interference visibility with
detector jitter taken into account\cite{legero_2004}. 
$\mathcal{C}(\Delta t)$ represents the coincidence probability at time interval $\Delta t$ while 
$C_0$ is the baseline coincidence. Eqs~(6A-6C) show all coincidence probabilities are phase-dependent. $\mathcal{C}_0$ and $\mathcal{C}(\pm \tau)$'s dependence arises from the first-order interference, while $\mathcal{C}(0)$ contains contributions from  $|2\rangle_t$ states as well as incomplete HOM interference between two RF photons emitted separately by the AMZI delay $\tau$.  Figures~\ref{fig4}\textbf{b}, 4\textbf{c} plot the phase dependence of the theoretical (solid lines) and experimental (symbols) coincidence rates for $\Delta t = \pm \tau$ and $\Delta t = 0$, normalised to the baseline coincidence. We use maximum likelihood estimation method to determine a realistic set of parameters for each excitation flux that provide the best fit to the data.  The theoretical simulations are in excellent agreement with the experimental data for three incident fluxes and have also successfully reproduced the crossover between $\mathcal{C}(\pm\tau)$ and $\mathcal{C}_0$ for $\bar{n} = 0.25$.
\vspace{0.3 cm}

\noindent\textbf{\large Discussion}\\ 
Over past 50 years, it has been prevalent to discuss resonance fluorescence in the context of ``coherently" and ``incoherently" scattered light~\cite{mollow1969,Lopez_Carreno_2018,phillips2020,Hanschke2020,masters2023,Konthasinghe2012, casalengua2024two,nguyen2011ultra}. In literature, interchangeable terminologies, such as resonant Rayleigh scattering (RRS) vs. resonant photoluminescence (RPL)~\cite{Proux2015} and elastic vs. inelastic scattering~\cite{metcalfe2013}, are also in use.
The term ``incoherent scattering" is rather misleading.  As we have elucidated,  both the laser-like and broadband parts arise from the very same coherent process, i.e., resonant absorption and spontaneous emission.  The two parts are integral. Their integrity is key to the joint observation of ``sub-natural" linewidth and anti-bunching.  Conversely, compromise in the integrity will change the photonic state and may lead to different observations, e.g., loss of anti-bunching after spectral filtering~\cite{Hanschke2020,phillips2020,masters2023} or super-bunching after the AMZI filtering (Fig.~\ref{fig3}). We stress that photon bunching does not necessarily require simultaneous scattering of two photons~\cite{masters2023} for an explanation. 

To conclude, we have presented a \zy{unified} model to explain the coherence of resonance fluorescence under continuous-wave excitation.  
It links the RF's coherence to the incident flux down to the single-photon level and we show how to manipulate photon number statistics through simple two-path interference. We clarify that coherent scattering can be treated as a process of absorption and re-emission and does not need to involve higher-order scattering processes~\cite{masters2023} to explain experiments. Our work adds clarity to the knowledge pool of RF-based quantum light sources and we believe it will help foster new applications. 
One opportunity is to exploit the RF's coherence for quantum secure communication~\cite{zhou2022b,karli_2024}.

\vspace{0.3 cm}
{\color{black}
\begin{figure}[htbp]
\centering
\includegraphics[width=1\columnwidth]{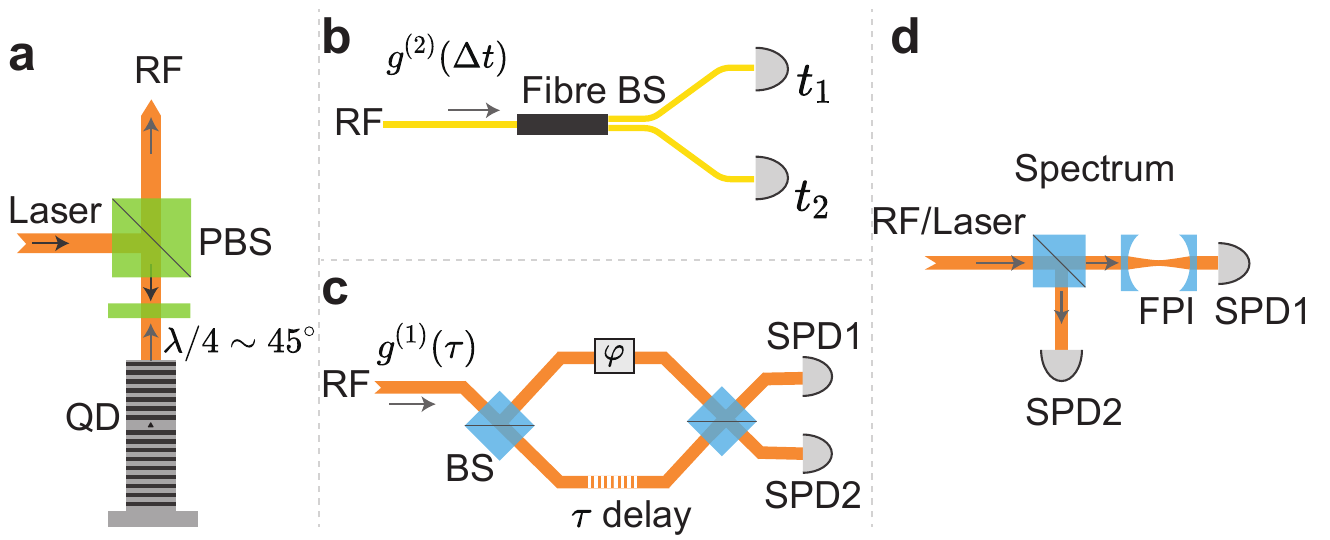}
\caption{\textbf{Experimental setup.} \textbf{a}, Confocal RF; \textbf{b}, HBT;  \textbf{c}, AMZI for $g^{(1)}(\tau)$ measurement; \textbf{d}, High-resolution spectral measurement.}
	\label{setup}
\end{figure}

\begin{figure}[htbp]
\includegraphics[width=.8\columnwidth]{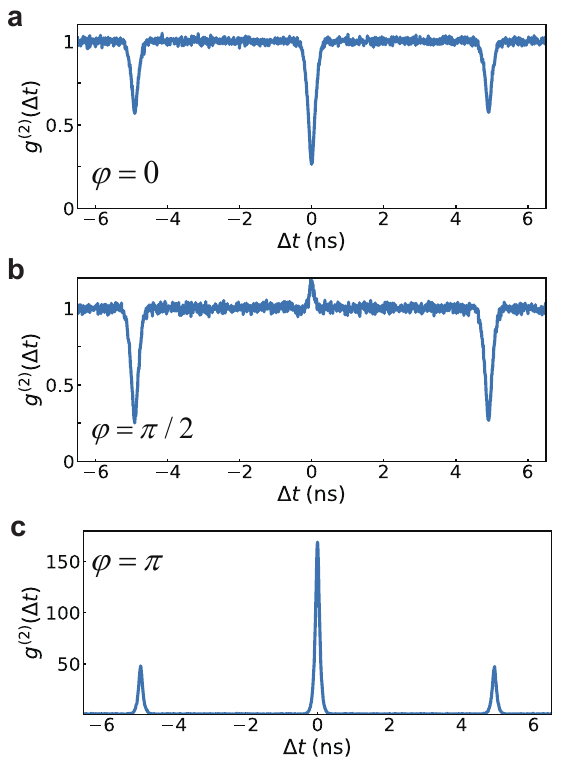}
\caption{\textbf{Phase-dependent auto-correlation of the AMZI-filtered RF.}  The excitation flux is set at $\bar{n} = 0.0062$.
 \textbf{a}, Setup; \textbf{b}, $\varphi = 0$; 
 \textbf{c}, $\varphi=\pi/2$; \textbf{d}, $\varphi = \pi$.}
	\label{fig:AMZI-HBT}
\end{figure}

\noindent\textbf{{\large Methods}}\\
\textbf{Experimental setup}\\
The main RF setup %experimental apparatus 
is shown in  Fig.~\ref{setup}\textbf{a}. Here,  a polarising beam splitter (PBS) and a $\sim$45$^\circ$ quarter-wave plate are used together as an optical router to direct the RF from the QD to the measurement apparatuses shown in panels \textbf{b}, \textbf{c} and \textbf{d}.
A CW laser (M SQUARED SolsTis PSX XF 5000, $\sim$100~kHz linewidth) is used as the excitation source.  
Unlike typical RF setups~\cite{Proux2015,PhysRevLett.100.207405}, the reflected laser and the RF 
 in our experiment have the same polarisation, thanks to our microcavity~\cite{wu2023} that suppresses the laser reflection.
 Fig.~\ref{setup}\textbf{b} shows a standard Hanbury Brown-Twiss (HBT) setup for measuring the auto-correlation function 
$g^{(2)}(\Delta t)$ that evaluates the single-photon purity of the input signal. It consists of a 50:50 fibre beam splitter and two single photon detectors. %An ideal single-photon state corresponds to $g^{(2)}(0) = 0$.}

Fig.~\ref{setup}\textbf{c} illustrates a setup for characterising the first-order correlation function $g^{(1)}(\tau)$. In this setup, both beam splitters have a nominal 50:50 reflectance-to-transmittance ratio, and the AMZI's differential delay is 4.92~ns. The count rates at the detectors oscillate with a free-drifting phase $\varphi$. By measuring the maximum and minimum values of this oscillation, we can calculate the interference fringe visibility: $V \equiv |g^{(1)}(\tau)| = \frac{C_{max} - C_{min}}{C_{max}+C_{min}}$. Usually, one detector would suffice. However, to avoid the QD blinking affecting the measurement result, we use a two-channel summation method to normalise each detector's count rate to the combined count rate for the visibility calculation. 

In measuring the high resolution RF spectra, we use a setup shown 
in Fig.~\ref{setup}\textbf{d}. The signal is split into two paths. One path enters the scanning FPI with a single photon detector (SPD1) recording the signal count rate as a function of the FPI transmission frequency  which is controlled by a piezo actuator. The other path enters a second single photon detector (SPD2) for normalising SPD1's detection results. The scanning FPI has a free spectral range of 20~GHz and a resolution of 20~MHz.

Two superconducting nanowire single photon detectors (SNSPDs) are used and characterised to have a detection efficiency of 78~\% and a time jitter of 48~ps at a wavelength of 910~nm. A time-tagger is used for correlation and time-resolved measurements.

\vspace{0.3 cm}
\noindent\textbf{Phase-locking the AMZI}\\
\noindent  
The AMZI is made from optical fibres and sealed in an enclosure to shield from external airflow and ambient temperature fluctuation. However, its phase still drifts on the order of $\pi$ per minute. While adequate for the first-order coherence measurement (e.g., see Fig.~1\textbf{c}), this level of stability does not meet the requirement by our phase-dependent correlation experiments. To eliminate the phase drift,
we use a thermo-electric cooler to actively control the internal temperature of the enclosure, based on the ratio of the single-photon counting rates between the AMZI outputs.

While having been demonstrated in Figs.~\ref{fig3} and \ref{fig4}, the effectiveness of the phase-locking can be further appreciated from the phase-dependent auto-correlation data shown in Fig.~\ref{fig:AMZI-HBT}. Using the setup shown in Fig.~\ref{fig3}\textbf{a}, the photon correlation statistics of the AMZI-filtered RF signal can be tuned from anti-bunching ($g^{(2)}(0) < 1$) at $\varphi =0$ to super-bunching ($g^{(2)}(0) \gg 1$) at $\varphi = \pi$. The result is consistent with our theoretical calculation, where we obtain $g^{(2)}(0) \equiv \frac{\mathcal{C}(0)}{\mathcal{C}_0} = \frac{1}{(1+p_0 \cos \varphi)^2}$ with $\mathcal{C}(0)=p_1^2/16$ and $\mathcal{C}_0=(1+p_0 \cos\varphi)^2p_1^2/16$ being 
the 0-delay and baseline coincidence probabilities, respectively.}

\vspace{0.3 cm}
\noindent\textbf{\large Data availability}\\
All the data that support the plots within this manuscript and other findings of this study are available from Z.Y. upon reasonable request.

%\nocite{*}

\begin{thebibliography}{10}
\expandafter\ifx\csname url\endcsname\relax
  \def\url#1{\texttt{#1}}\fi
\expandafter\ifx\csname urlprefix\endcsname\relax\def\urlprefix{URL }\fi
\providecommand{\bibinfo}[2]{#2}
\providecommand{\eprint}[2][]{\url{#2}}

\bibitem{scully_zubairy_1997}
\bibinfo{author}{Scully, M.~O.} \& \bibinfo{author}{Zubairy, M.~S.}
\newblock \emph{\bibinfo{title}{Quantum Optics}} (\bibinfo{publisher}{Cambridge University Press}, \bibinfo{year}{1997}).

\bibitem{mandel1995optical}
\bibinfo{author}{Mandel, L.} \& \bibinfo{author}{Wolf, E.}
\newblock \emph{\bibinfo{title}{Optical coherence and quantum optics}} (\bibinfo{publisher}{Cambridge university press}, \bibinfo{year}{1995}).

\bibitem{loudon2000quantum}
\bibinfo{author}{Loudon, R.}
\newblock \emph{\bibinfo{title}{The quantum theory of light}} (\bibinfo{publisher}{OUP Oxford}, \bibinfo{year}{2000}).

\bibitem{Steck2023}
\bibinfo{author}{Steck, D.~A.}
\newblock \bibinfo{title}{Quantum and atom optics}.
\newblock \bibinfo{note}{\url{http://steck.us/teaching (revision 0.14, 23 August 2023)}.}

\bibitem{Lopez_Carreno_2018}
\bibinfo{author}{{López~Carreño}, J.~C.}, \bibinfo{author}{{Zubizarreta~Casalengua}, E.}, \bibinfo{author}{Laussy, F.~P.} \& \bibinfo{author}{{del Valle}, E.}
\newblock \bibinfo{title}{Joint subnatural-linewidth and single-photon emission from resonance fluorescence}.
\newblock \emph{\bibinfo{journal}{Quant. Sci. Technol.}} \textbf{\bibinfo{volume}{3}}, \bibinfo{pages}{045001} (\bibinfo{year}{2018}).

\bibitem{phillips2020}
\bibinfo{author}{Phillips, C.~L.} \emph{et~al.}
\newblock \bibinfo{title}{Photon statistics of filtered resonance fluorescence}.
\newblock \emph{\bibinfo{journal}{Phys. Rev. Lett.}} \textbf{\bibinfo{volume}{125}}, \bibinfo{pages}{043603} (\bibinfo{year}{2020}).

\bibitem{Hanschke2020}
\bibinfo{author}{Hanschke, L.} \emph{et~al.}
\newblock \bibinfo{title}{Origin of antibunching in resonance fluorescence}.
\newblock \emph{\bibinfo{journal}{Phys. Rev. Lett.}} \textbf{\bibinfo{volume}{125}}, \bibinfo{pages}{170402} (\bibinfo{year}{2020}).

\bibitem{masters2023}
\bibinfo{author}{Masters, L.} \emph{et~al.}
\newblock \bibinfo{title}{On the simultaneous scattering of two photons by a single two-level atom}.
\newblock \emph{\bibinfo{journal}{Nat. Photon.}} \textbf{\bibinfo{volume}{17}}, \bibinfo{pages}{972--976} (\bibinfo{year}{2023}).

\bibitem{casalengua2024two}
\bibinfo{author}{Casalengua, E.~Z.}, \bibinfo{author}{Laussy, F.~P.} \& \bibinfo{author}{del Valle, E.}
\newblock \bibinfo{title}{Two photons everywhere}.
\newblock \emph{\bibinfo{journal}{Phil. Trans. R. Soc. A}} \textbf{\bibinfo{volume}{382}}, \bibinfo{pages}{20230315} (\bibinfo{year}{2024}).

\bibitem{Ng_2022}
\bibinfo{author}{Ng, B.~L.}, \bibinfo{author}{Chow, C.~H.} \& \bibinfo{author}{Kurtsiefer, C.}
\newblock \bibinfo{title}{Observation of the {Mollow} triplet from an optically confined single atom}.
\newblock \emph{\bibinfo{journal}{Phys. Rev. A}} \textbf{\bibinfo{volume}{106}}, \bibinfo{pages}{063719} (\bibinfo{year}{2022}).

\bibitem{wu2023}
\bibinfo{author}{Wu, B.} \emph{et~al.}
\newblock \bibinfo{title}{Mollow triplets under few-photon excitation}.
\newblock \emph{\bibinfo{journal}{Optica}} \textbf{\bibinfo{volume}{10}}, \bibinfo{pages}{1118 -- 1123} (\bibinfo{year}{2023}).

\bibitem{hoffges1997heterodyne}
\bibinfo{author}{H{\"o}ffges, J.}, \bibinfo{author}{Baldauf, H.}, \bibinfo{author}{Eichler, T.}, \bibinfo{author}{Helmfrid, S.} \& \bibinfo{author}{Walther, H.}
\newblock \bibinfo{title}{Heterodyne measurement of the fluorescent radiation of a single trapped ion}.
\newblock \emph{\bibinfo{journal}{Opt. Commun.}} \textbf{\bibinfo{volume}{133}}, \bibinfo{pages}{170--174} (\bibinfo{year}{1997}).

\bibitem{nguyen2011ultra}
\bibinfo{author}{Nguyen, H.-S.} \emph{et~al.}
\newblock \bibinfo{title}{Ultra-coherent single photon source}.
\newblock \emph{\bibinfo{journal}{Appl. Phys. Lett.}} \textbf{\bibinfo{volume}{99}}, \bibinfo{pages}{261904} (\bibinfo{year}{2011}).

\bibitem{Konthasinghe2012}
\bibinfo{author}{Konthasinghe, K.} \emph{et~al.}
\newblock \bibinfo{title}{Coherent versus incoherent light scattering from a quantum dot}.
\newblock \emph{\bibinfo{journal}{Phys. Rev. B}} \textbf{\bibinfo{volume}{85}}, \bibinfo{pages}{235315} (\bibinfo{year}{2012}).

\bibitem{mattiesen2012}
\bibinfo{author}{Matthiesen, C.}, \bibinfo{author}{Vamivakas, A.~N.} \& \bibinfo{author}{Atat\"ure, M.}
\newblock \bibinfo{title}{Subnatural linewidth single photons from a quantum dot}.
\newblock \emph{\bibinfo{journal}{Phys. Rev. Lett.}} \textbf{\bibinfo{volume}{108}}, \bibinfo{pages}{093602} (\bibinfo{year}{2012}).

\bibitem{matthiesen2013phase}
\bibinfo{author}{Matthiesen, C.} \emph{et~al.}
\newblock \bibinfo{title}{Phase-locked indistinguishable photons with synthesized waveforms from a solid-state source}.
\newblock \emph{\bibinfo{journal}{Nat. Commun.}} \textbf{\bibinfo{volume}{4}}, \bibinfo{pages}{1600} (\bibinfo{year}{2013}).

\bibitem{Carmichael_1976}
\bibinfo{author}{Carmichael, H.~J.} \& \bibinfo{author}{Walls, D.~F.}
\newblock \bibinfo{title}{Proposal for the measurement of the resonant stark effect by photon correlation techniques}.
\newblock \emph{\bibinfo{journal}{J. Phys. B At. Mol. Opt. Phys.}} \textbf{\bibinfo{volume}{9}}, \bibinfo{pages}{L43} (\bibinfo{year}{1976}).

\bibitem{kimble1977}
\bibinfo{author}{Kimble, H.~J.}, \bibinfo{author}{Dagenais, M.} \& \bibinfo{author}{Mandel, L.}
\newblock \bibinfo{title}{Photon antibunching in resonance fluorescence}.
\newblock \emph{\bibinfo{journal}{Phys. Rev. Lett.}} \textbf{\bibinfo{volume}{39}}, \bibinfo{pages}{691--695} (\bibinfo{year}{1977}).

\bibitem{lodahl2015}
\bibinfo{author}{Lodahl, P.}, \bibinfo{author}{Mahmoodian, S.} \& \bibinfo{author}{Stobbe, S.}
\newblock \bibinfo{title}{Interfacing single photons and single quantum dots with photonic nanostructures}.
\newblock \emph{\bibinfo{journal}{Rev. Mod. Phys.}} \textbf{\bibinfo{volume}{87}}, \bibinfo{pages}{347--400} (\bibinfo{year}{2015}).

\bibitem{mollow1969}
\bibinfo{author}{Mollow, B.~R.}
\newblock \bibinfo{title}{Power spectrum of light scattered by two-level systems}.
\newblock \emph{\bibinfo{journal}{Phys. Rev.}} \textbf{\bibinfo{volume}{188}}, \bibinfo{pages}{1969--1975} (\bibinfo{year}{1969}).

\bibitem{Hong1987}
\bibinfo{author}{Hong, C.~K.}, \bibinfo{author}{Ou, Z.~Y.} \& \bibinfo{author}{Mandel, L.}
\newblock \bibinfo{title}{Measurement of subpicosecond time intervals between two photons by interference}.
\newblock \emph{\bibinfo{journal}{Phys. Rev. Lett.}} \textbf{\bibinfo{volume}{59}}, \bibinfo{pages}{2044--2046} (\bibinfo{year}{1987}).

\bibitem{Wrigge2008}
\bibinfo{author}{Wriggle, G.}, \bibinfo{author}{Gerhardt, I.}, \bibinfo{author}{Hwang, J.}, \bibinfo{author}{Zumofen, G.} \& \bibinfo{author}{Sandoghdar, V.}
\newblock \bibinfo{title}{Efficient coupling of photons to a single molecule and the observation of its resonance fluorescence}.
\newblock \emph{\bibinfo{journal}{Nat. Phys.}} \textbf{\bibinfo{volume}{4}}, \bibinfo{pages}{60--66} (\bibinfo{year}{2008}).

\bibitem{ates2009}
\bibinfo{author}{Ates, S.} \emph{et~al.}
\newblock \bibinfo{title}{Post-{Selected} {Indistinguishable} {Photons} from the {Resonance} {Fluorescence} of a {Single} {Quantum} {Dot} in a {Microcavity}}.
\newblock \emph{\bibinfo{journal}{Phys. Rev. Lett.}} \textbf{\bibinfo{volume}{103}}, \bibinfo{pages}{167402} (\bibinfo{year}{2009}).

\bibitem{flagg_2009}
\bibinfo{author}{Flagg, E.~B.} \emph{et~al.}
\newblock \bibinfo{title}{Resonantly driven coherent oscillations in a solid-state quantum emitter}.
\newblock \emph{\bibinfo{journal}{Nat. Phys.}} \textbf{\bibinfo{volume}{5}}, \bibinfo{pages}{203--207} (\bibinfo{year}{2009}).

\bibitem{de_santis_2017}
\bibinfo{author}{De~Santis, L.} \emph{et~al.}
\newblock \bibinfo{title}{A solid-state single-photon filter}.
\newblock \emph{\bibinfo{journal}{Nat. Nanotech.}} \textbf{\bibinfo{volume}{12}}, \bibinfo{pages}{663--667} (\bibinfo{year}{2017}).

\bibitem{Proux2015}
\bibinfo{author}{Proux, R.} \emph{et~al.}
\newblock \bibinfo{title}{Measuring the photon coalescence time window in the continuous-wave regime for resonantly driven semiconductor quantum dots}.
\newblock \emph{\bibinfo{journal}{Phys. Rev. Lett.}} \textbf{\bibinfo{volume}{114}}, \bibinfo{pages}{067401} (\bibinfo{year}{2015}).

\bibitem{incoherent}
\bibinfo{note}{Consider an incoherently pumped TLE with a classical probability of $p_1$ emitting a photon. The HBT coincidence probability at $\Delta t = 0$ is proportional to $(p_1/2)^2 \times 1/2 \times 1/2 = p_1^2/16$, which is identical to the baseline coincidence probability at $\Delta t \neq 0, \pm \tau$.}

\bibitem{bennett2016semiconductor}
\bibinfo{author}{Bennett, A.} \emph{et~al.}
\newblock \bibinfo{title}{A semiconductor photon-sorter}.
\newblock \emph{\bibinfo{journal}{Nat. Nanotech.}} \textbf{\bibinfo{volume}{11}}, \bibinfo{pages}{857--860} (\bibinfo{year}{2016}).

\bibitem{loredo2019generation}
\bibinfo{author}{Loredo, J.} \emph{et~al.}
\newblock \bibinfo{title}{Generation of non-classical light in a photon-number superposition}.
\newblock \emph{\bibinfo{journal}{Nat. Photon.}} \textbf{\bibinfo{volume}{13}}, \bibinfo{pages}{803--808} (\bibinfo{year}{2019}).

\bibitem{iles-smith_2017}
\bibinfo{author}{Iles-Smith, J.}, \bibinfo{author}{McCutcheon, D. P.~S.}, \bibinfo{author}{Nazir, A.} \& \bibinfo{author}{Mørk, J.}
\newblock \bibinfo{title}{Phonon scattering inhibits simultaneous near-unity efficiency and indistinguishability in semiconductor single-photon sources}.
\newblock \emph{\bibinfo{journal}{Nat. Photon.}} \textbf{\bibinfo{volume}{11}}, \bibinfo{pages}{521--526} (\bibinfo{year}{2017}).

\bibitem{koong2019}
\bibinfo{author}{Koong, Z.~X.} \emph{et~al.}
\newblock \bibinfo{title}{Fundamental limits to coherent photon generation with solid-state atomlike transitions}.
\newblock \emph{\bibinfo{journal}{Phys. Rev. Lett.}} \textbf{\bibinfo{volume}{123}}, \bibinfo{pages}{167402} (\bibinfo{year}{2019}).

\bibitem{brash2019}
\bibinfo{author}{Brash, A.~J.} \emph{et~al.}
\newblock \bibinfo{title}{Light scattering from solid-state quantum emitters: Beyond the atomic picture}.
\newblock \emph{\bibinfo{journal}{Phys. Rev. Lett.}} \textbf{\bibinfo{volume}{123}}, \bibinfo{pages}{167403} (\bibinfo{year}{2019}).

\bibitem{zhai2020}
\bibinfo{author}{Zhai, L.} \emph{et~al.}
\newblock \bibinfo{title}{Low-noise {GaAs} quantum dots for quantum photonics}.
\newblock \emph{\bibinfo{journal}{Nat. Commun.}} \textbf{\bibinfo{volume}{11}}, \bibinfo{pages}{4745} (\bibinfo{year}{2020}).

\bibitem{PhysRevLett.100.207405}
\bibinfo{author}{Patel, R.~B.} \emph{et~al.}
\newblock \bibinfo{title}{Postselective two-photon interference from a continuous nonclassical stream of photons emitted by a quantum dot}.
\newblock \emph{\bibinfo{journal}{Phys. Rev. Lett.}} \textbf{\bibinfo{volume}{100}}, \bibinfo{pages}{207405} (\bibinfo{year}{2008}).

\bibitem{legero_2004}
\bibinfo{author}{Legero, T.}, \bibinfo{author}{Wilk, T.}, \bibinfo{author}{Hennrich, M.}, \bibinfo{author}{Rempe, G.} \& \bibinfo{author}{Kuhn, A.}
\newblock \bibinfo{title}{Quantum beat of two single photons}.
\newblock \emph{\bibinfo{journal}{Phys. Rev. Lett.}} \textbf{\bibinfo{volume}{93}}, \bibinfo{pages}{070503} (\bibinfo{year}{2004}).

\bibitem{metcalfe2013}
\bibinfo{author}{Metcalfe, M.}, \bibinfo{author}{Solomon, G.~S.} \& \bibinfo{author}{Lawall, J.}
\newblock \bibinfo{title}{{Heterodyne measurement of resonant elastic scattering from epitaxial quantum dots}}.
\newblock \emph{\bibinfo{journal}{Appl. Phys. Lett.}} \textbf{\bibinfo{volume}{102}}, \bibinfo{pages}{231114} (\bibinfo{year}{2013}).

\bibitem{zhou2022b}
\bibinfo{author}{Zhou, L.} \emph{et~al.}
\newblock \bibinfo{title}{Experimental quantum communication overcomes the rate-loss limit without optical phase tracking}.
\newblock \emph{\bibinfo{journal}{Phys. Rev. Lett.}} \textbf{\bibinfo{volume}{130}}, \bibinfo{pages}{250801} (\bibinfo{year}{2023}).

\bibitem{karli_2024}
\bibinfo{author}{Karli, Y.} \emph{et~al.}
\newblock \bibinfo{title}{Controlling the photon number coherence of solid-state quantum light sources for quantum cryptography}.
\newblock \emph{\bibinfo{journal}{npj Quant. Inf.}} \textbf{\bibinfo{volume}{10}}, \bibinfo{pages}{17} (\bibinfo{year}{2024}).

\end{thebibliography}

~

% \noindent\textbf{Data availability}\\
% All the data that support the plots within this manuscript and other findings of this study are available from Z.Y. upon reasonable request.

~

\noindent\textbf{Acknowledgements}\\ 
The authors thank Z.~Q. Yin, Y.~K.~Wu, Y. Ji, X. B. Wang and C. Antón-Solanas for helpful discussions, and S. Wein for help on  derivation of Eq.~(1).
This work was supported by Beijing Natural Science Foundation under grant IS23011,
the National Natural Science
Foundation of China under grants 12204049, 12274223,
62250710162, 12494600 and 12494604, and
National Key R \& D Program of China under grant 2018YFA0306101.

~ 

\noindent\textbf{Author contributions}\\
Z.Y., X.-J.W. and B.W. designed the research.  X.-J.W., G.H. and B.W. carried out the experiments.  H.-L.Y., M.-Y.L., Y.-Z.W., and X.-J.W. developed the theoretical derivation and performed the simulations. L.L. fabricated the devices with assistance from W.J. H.L., H.N. and Z.N. grew the semiconductor wafer. X.-J.W. and Z.Y. prepared the manuscript with input from all authors.  Z.Y. conceived the model and supervised the project.

~

\noindent\textbf{Competing interests}\\
The authors declare no competing interests. \\

%\noindent\textbf{Additional Information}\\

\noindent  \textbf{Correspondence} and requests for materials should be addressed to Zhiliang Yuan. 
\end{document}

\title{Supplementary Information for ``Coherence in Resonance Fluorescence"}

\author{Xu-Jie~Wang}
\affiliation{Beijing Academy of Quantum Information Sciences, Beijing 100193, China}
\author{Guoqi~Huang}
\affiliation{Beijing Academy of Quantum Information Sciences, Beijing 100193, China}
\affiliation{School of Science, Beijing University of Posts and Telecommunications, Beijing 100876, China}
\author{Ming-Yang~Li}
\author{Yuan-Zhuo~Wang}
\affiliation{National Laboratory of Solid State Microstructures and School of Physics,Collaborative Innovation Center of Advanced Microstructures, Nanjing University, Nanjing 210093, China}
\author{Li~Liu}
\affiliation{Beijing Academy of Quantum Information Sciences, Beijing 100193, China}
\author{Bang~Wu}
\email{wubang@baqis.ac.cn}
\affiliation{Beijing Academy of Quantum Information Sciences, Beijing 100193, China}
\author{Hanqing~Liu}
\author{Haiqiao~Ni}
\author{Zhichuan~Niu}
\affiliation{Key Laboratory of Optoelectronic Materials and Devices, Institute of Semiconductors, Chinese Academy of Sciences, Beijing 100083, China}
%\affiliation{State Key Laboratory of Superlattices and Microstructures, Institute of Semiconductors, Chinese Academy of Sciences, Beijing 100083, China}
\affiliation{Center of Materials Science and Optoelectronics Engineering, University of Chinese Academy of Sciences, Beijing 100049, China}
\author{Weijie~Ji}
\affiliation{Beijing Academy of Quantum Information Sciences, Beijing 100193, China}
\author{Rongzhen~Jiao}
\affiliation{School of Science, Beijing University of Posts and Telecommunications, Beijing 100876, China}
\author{Hua-Lei~Yin}
\email{hlyin@ruc.edu.cn}
\affiliation{Beijing Academy of Quantum Information Sciences, Beijing 100193, China}
\affiliation{National Laboratory of Solid State Microstructures and School of Physics,Collaborative 
Innovation Center of Advanced Microstructures, Nanjing University, Nanjing 210093, China}
\affiliation{School of Physics and Beijing Key Laboratory of Opto-electronic Functional Materials and Micro-nano Devices, Key Laboratory of Quantum State Construction and Manipulation (Ministry of Education), Renmin University of China, Beijing 100872, China}
\author{Zhiliang~Yuan}%
\email{yuanzl@baqis.ac.cn}
\affiliation{Beijing Academy of Quantum Information Sciences, Beijing 100193, China}

\date{\today}

\maketitle
{\color{black}
\section{Pure-state description of resonance fluorescence of a two-level emitter}

In theoretical treatments for resonance fluorescence (RF) of a two-level emitter (TLE), it is common for the TLE and the fluorescence photon to be considered separately, and either of the separate sub-systems is a mixed state~\cite{scully_zubairy_1997,karli_2024}.  In Main Text, we propose a pure state model  that treats the TLE and the fluorescence photon jointly, and have found the pure state  useful in guiding the experiments. Here, we present a derivation of the pure-state using the quantum optics master equation theory.  

\subsection{Density matrices} 

For a TLE that is coherently driven by a laser field, density matrices can be used to describe the TLE state, the spontaneous emission photon (photon) state, or the joint TLE-photon state. If we assume that there is no multi-excitation component, then the instantaneous joint TLE-photon state can be written as
\begin{equation}
		\hat{\rho}_{ap}=\left(\begin{array}{lll}
			\rho_{g 0, g 0} & \rho_{g 0, g 1} & \rho_{g 0, e 0} \\
			\rho_{g 1, g 0} & \rho_{g 1, g 1} & \rho_{g 1, e 0} \\
			\rho_{e 0, g 0} & \rho_{e 0, g 1} & \rho_{e 0, e 0}
		\end{array}\right),
\end{equation}
where $\ket{e}$ and $\ket{g}$ are the 
excited state and ground state of the TLE while $\ket{0}$ and $\ket{1}$ are vacuum state and single-photon state of resonance fluorescence.
$\rho_{ab,cd}$ is the matrix element for $\ket{a}\bra{b}\otimes\ket{c}\bra{d}$. According to the definition of a valid density matrix, we have $\rho_{g0,g0}+\rho_{g1,g1}+\rho_{e0, e0}=1$, $\rho_{g0,g1}=\rho_{g1,g0}^{*}$, $\rho_{g0,e0}=\rho_{e0,g0}^{*}$ and $\rho_{g1,e0}=\rho_{e0,g1}^{*}$. \zy{
The assumption of no multi-excitations is justifiable, as ideally a coherently driven TLE never emits two photons simultaneously~\cite{scully_zubairy_1997}.}

%The reduced density matrices for the matter  ($\rho$) and the photon ($\rho_p$) sub-system can be obtained from $\hat{\rho}_{ap}$ by performing partial traces, i.e.
%\begin{equation}
%\rho=\text{tr}_p(\rho_{ap})=\left(\begin{array}{cc}
%\rho_{g 0, g 0}+\rho_{g 1, g 1}  & \rho_{g 0, e 0} \\
 %\rho_{e0, g0} & \rho_{e0, e0} 
%\end{array}\right),
%\end{equation}
%\begin{equation}
%\rho_{p}=\text{tr}_a(\rho_{ap})=\left(\begin{array}{cc}
%\rho_{g 0, g 0} +\rho_{e 0,e 0} & \rho_{g 0, g 1} \\
%\rho_{g 1, g 0} & \rho_{g 1, g 1}  
%\end{array}\right).
%\end{equation}

The reduced density matrix of the TLE sub-system can be expressed as $\hat{\rho}_{a}={\rm Tr}_{p}[\hat{\rho}_{ap}]=\rho_{gg}\ket{g}\bra{g}+\rho_{ge}\ket{g}\bra{e}+\rho_{eg}\ket{e}\bra{g}+\rho_{ee}\ket{e}\bra{e}$, where the matrix elements are related to those in $\hat{\rho}_{ap}$:
\begin{equation}
\label{eq2}
\begin{aligned}
\rho_{gg}&=\rho_{g0,g0}+\rho_{g1,g1},~~~~\rho_{ge}=\rho_{g0,e0},\\
\rho_{eg}&=\rho_{e0,g0},~~~~~~~~~~~~~~~\rho_{ee}=\rho_{e0,e0}.\\
\end{aligned}
\end{equation}
Similarly, the reduced density matrix of the \zy{\textit{instantaneous}} photon sub-system can be expressed as $\hat{\rho}_{p}={\rm Tr}_{a}[\hat{\rho}_{ap}]=\rho_{00}\ket{0}\bra{0}+\rho_{01}\ket{0}\bra{1}+\rho_{10}\ket{1}\bra{0}+\rho_{11}\ket{1}\bra{1}$, where we have the matrix elements
\begin{equation}
\begin{aligned}\label{eq3}
\rho_{00}&=\rho_{g0,g0}+\rho_{e0,e0},~~~~\rho_{01}=\rho_{g0,g1},\\
\rho_{10}&=\rho_{g1,g0},~~~~~~~~~~~~~~~\rho_{11}=\rho_{g1,g1}.\\
\end{aligned}
\end{equation}

%Specially, if there is no pure dephasing effect, the joint quantum state of the two-level system and the spontaneous emission photon system will be a pure state, which is a main conclusion in Main Text and will be proven in the following.

\subsection{The TLE sub-system}

We first review a textbook treatment for spontaneous emission under continuous-wave driving~\cite{scully_zubairy_1997} and then obtain the density matrix of the TLE in the case of the steady-state.

Consider a two-level system model in the frame rotating with the laser, and with a continuous-wave driving form. The total Hamiltonian in the interaction picture can be expressed as
\begin{equation}
\begin{aligned}\label{}
H_{I}=-\frac{\hbar\Omega }{2}(e^{-\textbf{i}\phi}\sigma_{+}+e^{\textbf{i}\phi}\sigma_{-}),
\end{aligned}
\end{equation}
where the excitation field is treated using semi-classical theory, $\Omega$ is defined as the classical Rabi frequency and $\phi$ is the phase of the dipole matrix element $\mathbf{p}_{ge}=|\mathbf{p}_{ge}|e^{i\phi}$. 
$\sigma_{+}=\ket{e}\bra{g}$ and $\sigma_{-}=\ket{g}\bra{e}$ are the the raising and lowering operators of a two-level system, and the Pauli operator $\sigma_{z}$ used later can be written as $\sigma_{z}=\ket{e}\bra{e}-\ket{g}\bra{g}$ accordingly. Since the phase $\phi$ has no observable effects, we can let  $\phi=\frac{\pi}{2}$ and simplify the Hamiltonian to
\begin{equation}
\begin{aligned}\label{}
H_{I}=\textbf{i}\frac{\hbar\Omega }{2}(\sigma_{+}-\sigma_{-}).
\end{aligned}
\end{equation}

Taking into account spontaneous emission under the Markov approximation, the dynamics of the two-level system is ruled by the Lindbladt Master equation%~\cite{scully_zubairy_1997}
\begin{equation}
\begin{aligned}\label{eq6}
\dot{\hat{\rho}}_{a}=-\frac{\textbf{i}}{\hbar}[H_{I},\rho_{a}]+D_{\gamma,\sigma_{-}}[\rho_{a}]+D_{\gamma^{*}/2,\sigma_{z}}[\rho_{a}],
\end{aligned}
\end{equation}
where $\gamma=1/T_{1}$ and $\gamma^{*}$ are the spontaneous emission rate and pure dephasing rate, respectively,  and
$D_{x,A}[\rho]$ is the super-operators defined as $D_{x,A}[\rho]=\frac{x}{2}(2A\rho A^{\dagger}-A^{\dagger}A\rho-\rho A^{\dagger}A)$.
Starting with an initial state $\hat{\rho}_{a}(0)=\ket{g}\bra{g}$, the steady-state solution of the Eq.~\eqref{eq6} can be written as
	\begin{widetext}
\begin{equation}
\begin{aligned}\label{}
\hat{\rho}_{a}(t\rightarrow\infty)=\frac{1}{2(1+\Omega^{2}T_{1}T_{2})}\left[(2+\Omega^{2}T_{1}T_{2})\ket{g}\bra{g}+\Omega T_{2}\ket{g}\bra{e}+\Omega T_{2}\ket{e}\bra{g}+\Omega^{2} T_{1}T_{2}\ket{e}\bra{e}\right],
\end{aligned}
\end{equation}
	\end{widetext}
where $1/T_{2}=\frac{\gamma}{2}+\gamma^{*}$ is total decoherence rate and $T_{2}\leq2T_{1}$. We remark that the density matrix of the TLE sub-system $\hat{\rho}_{a}$ in the above equation is not a pure state but a mixed state.

\subsection{The photon sub-system}

Based on the Weisskopf-Winger approximation~\cite{scully_zubairy_1997}, the positive frequency part of the field operator is proportional to the two-level system lowering operator, i.e.,
\begin{equation}
\begin{aligned}\label{eq8}
{a}=B \sigma_{-},
\end{aligned}
\end{equation}
where $B$ is a coefficient.
It is precisely with Eq.~\eqref{eq8} that one can use the density matrix of the TLE sub-system to calculate the spectral properties  and the second-order correlation function of the spontaneous emission light field \zy{for a retarded time that depends on the distance between the emitter and the observation location}.
Similarly, we use the above relation to evaluate the coherence and intensity at some point time $t$ once the TLE is in the steady state:

\begin{widetext}
\begin{equation}
\begin{aligned}\label{eq9}
\langle a (t) \rangle _\infty &={\rm Tr}[\hat{a} \hat{\rho}_{p}]={\rm Tr}[\hat{a} \hat{\rho}_{ap}]={\rm Tr}[B\sigma_{-}\hat{\rho}_{ap}]={\rm Tr}[B\sigma_{-}\hat{\rho}_{a}]=B\rho_{ge}=B\frac{\Omega T_{2}}{2(1+\Omega^{2}T_{1}T_{2})},\\
\langle a ^\dagger (t) a (t) \rangle_\infty &={\rm Tr}[\hat{a}^\dagger \hat{a} \hat{\rho}_{p}]={\rm Tr}[\hat{a}^\dagger \hat{a} \hat{\rho}_{ap}]={\rm Tr}[B^{2}\sigma_{+}\sigma_{-}\hat{\rho}_{ap}]={\rm Tr}[B^{2}\sigma_{+}\sigma_{-}\hat{\rho}_{a}]=B^{2}\rho_{ee}=B^{2}\frac{\Omega^{2}T_{1}T_{2}}{2(1+\Omega^{2}T_{1}T_{2})}. 
\end{aligned}
\end{equation}
\end{widetext}
Normally, the square magnitude of the coherence is normalised by the intensity to get the first-order coherence, 
\begin{equation}\label{eq10}
    g^{(1)}_\infty = \frac{T_{2}/T_{1}}{2(1 +  \Omega^2T_{1}T_{2})}.
\end{equation}
\zy{This result is not new, and has been routinely used in simulating the excitation-power dependence of the laser-like spectral fraction in the total RF of a TLE~\cite{nguyen2011ultra,mattiesen2012,Proux2015, brash2019,koong2019}.}  We have $g^{(1)}_\infty \to 1$ when $\gamma^* \to 0$ and $\Omega \to 0$.

The exact value of the coefficient $B$ does not affect the coherence of the collected spontaneous emission field. This situation is similar to the conclusion of the next section, i.e., that channel losses do not alter coherence \zy{of a quantum mechanical state}.  Assuming that the proportionality of the dipole and field is valid,  the reduced \textit{instantaneous} photon state should be equal to the steady-state of the TLE, i.e., $B = 1$. We have
\begin{widetext}
\begin{equation}
\begin{aligned}\label{rho_a}
\hat{\rho}_{p}(t\rightarrow\infty)=\frac{1}{2(1+\Omega^{2}T_{1}T_{2})}\left[(2+\Omega^{2}T_{1}T_{2})\ket{0}\bra{0}+\Omega T_{2}\ket{0}\bra{1}+\Omega T_{2}\ket{1}\bra{0}+\Omega^{2} T_{1}T_{2}\ket{1}\bra{1}\right].
\end{aligned}
\end{equation}
\end{widetext}
We remark that the density matrix of the instantaneous photon sub-system $\hat{\rho}_{p}$ above is not a pure state but a mixed state.

\subsection{Density matrix of the joint TLE-photon system}

With the results of the reduced density matrices of the TLE and the photon sub-systems, we can determine the diagonal  elements of the density matrix of the joint TLE-photon system using Eqs.~\ref{eq2} and \ref{eq3}: %\xj{Based on Eq.~(\ref{rho_a}), we can determine 
\begin{equation}    
\begin{aligned}
\rho_{e0,e0}&=\rho_{ee}=\frac{\Omega^2T_1T_2}{2(1+\Omega^2T_1T_2)},\\
\rho_{g1,g1}&=\rho_{11}=\frac{\Omega^2T_1T_2}{2(1+\Omega^2T_1T_2)}, \\
  \rho_{g0,g0} &= 1-\rho_{g1,g1}-\rho_{e0,e0}=\frac{1}{1+\Omega^2T_1T_2}.
\end{aligned}
\end{equation}

The first-order coherence of the resonance fluorescence, $g^{(1)}_{\infty}$, can be calculated from either density matrix of the sub-systems, $\hat{\rho}_a$ or $\hat{\rho}_p$~\cite{karli_2024}. This leads to the relationship ${\left|\rho_{g e}\right|^2}/{\rho_{e e}}={\left|\rho_{01}\right|^2}/{\rho_{11}}$, from which we deduce 
\begin{equation}
\rho_{g 0, g 1}=\rho_{g 1, g 0}^*=\rho_{g 0, e 0}=\rho_{e 0, g 0}^*=\frac{\Omega T_2}{2(1+\Omega^2T_1T_2)}.
\end{equation}

Until now, we have the density matrix of the joint TLE-photon system in the following form:
\begin{equation}
		\hat{\rho}_{ap}=\frac{1}{2(1+\Omega^{2}T_{1}T_{2})}\left(\begin{array}{ccc}
			2& \Omega T_{2} & \Omega T_{2} \\
			\Omega T_{2} & \Omega^{2} T_{1}T_{2} & \rho_{g1,e0}^\prime \\
			\Omega T_{2} & \rho_{e0,g1}^\prime & \Omega^{2} T_{1}T_{2}
		\end{array}\right),
\end{equation}
where the elements $\rho_{g1,e0}^\prime$ and $\rho_{e0,g1}^\prime$ ($\rho_{g1,e0}^\prime = \rho^{\prime *}_{e0,g1}$) remain unconstrained.
This is not surprising because their values will not influence the state of the TLE when tracing out the photon, nor the state of the photon when tracing out the TLE. However, we do know that $\hat{\rho}_{ap}$ must be a valid density matrix, and so all of its eigen values must be non-negative for any $\Omega$, $T_1$ and $T_2$.  We can apply this constraint to infer the unknown matrix elements, as described below.

Let $\rho_{g1,e0}^\prime = x e^{\textbf{i}\theta}$, where $x\geq0$ and $\theta\in [0,2\pi)$.  
%However,  the value of $x$ must be chosen such that the density matrix $\hat{\rho}_{ap}$ is a positive semi-definite operator. The three eigenvalues of the density matrix $\hat{\rho}_{ap}$ should be greater than or equal to zero, \xj{thus 
The non-negative eigenvalues $\lambda_{1}$, $\lambda_{2}$ and $\lambda_{3}$ of the matrix $2(1+\Omega^{2}T_{1}T_{2}) \times \hat{\rho}_{ap}$ are the roots of the cubic equation of one variable as follows
\begin{equation}
\begin{aligned}\label{rho_a}
a\lambda^{3}+b\lambda^{2}+c\lambda+d=0,
\end{aligned}
\end{equation}
where we have
\begin{equation}
\begin{aligned}\label{rho_a}
\left\{\begin{array}{l}
a=1,\\
b=-2(1+\Omega^{2}T_{1}T_{2}),\\
c=-(x^{2}-4\Omega^{2}T_{1}T_{2}+\Omega^{2}T_{2}^{2}-\Omega^{4}T_{1}^{2}T_{2}^{2}),\\
d=2(x^{2}-x\Omega^{2}T_{2}^{2}\cos\theta-\Omega^{4}T_{1}^{2}T_{2}^{2}+\Omega^{4}T_{1}T_{2}^{3}).\\
\end{array}\right.
\end{aligned}
\end{equation}
Vieta's formulas relate the polynomial coefficients to signed sums of products of the roots as follows:
\begin{equation}
\begin{aligned}\label{rho_a}
\left\{\begin{array}{l}
\lambda_{1}+\lambda_{2}+\lambda_{3}=-\frac{b}{a},\\
\lambda_{1}\lambda_{2}+\lambda_{1}\lambda_{3}+\lambda_{2}\lambda_{3}=\frac{c}{a},\\
\lambda_{1}\lambda_{2}\lambda_{3}=-\frac{d}{a}.
\end{array}\right.
\end{aligned}
\end{equation}
Then, we obtain the inequalities below
% \begin{widetext}
\begin{equation}
\begin{aligned}\label{eq20}
\left\{\begin{array}{l}
x^2-4 T_1 T_2 \Omega^2+2 T_2^2 \Omega^2-T_1^2 T_2^2 \Omega^4 \leq 0 \\
x^2-x T_2^2 \Omega^2 \cos \theta+\left(T_2-T_1\right) T_1 T_2^2 \Omega^4 \leq 0.
\end{array}\right.
\end{aligned}
\end{equation}
% \end{widetext}
From the second inequality of Eq.~\eqref{eq20}, we can deduce
\begin{equation}
\begin{aligned}\label{}
\left\{\begin{array}{l}
x\geq\frac{T_{2}\Omega^{2}}{2}\left(T_{2}\cos\theta-\sqrt{T_{2}^{2}\cos^{2}\theta-4(T_{2}-T_{1})T_{1}}\right),\\
x\leq\frac{T_{2}\Omega^{2}}{2}\left(T_{2}\cos\theta+\sqrt{T_{2}^{2}\cos^{2}\theta-4(T_{2}-T_{1})T_{1}}\right)
\end{array}\right.\\
\end{aligned}
\end{equation}
As we define earlier $x\geq0$, we then have
\begin{equation}
\begin{aligned}\label{ineq1}
T_{2}\cos\theta+\sqrt{T_{2}^{2}\cos^{2}\theta-4(T_{2}-T_{1})T_{1}}\geq0,\\
\end{aligned}
\end{equation}
which requires 
\begin{equation}\label{}
T_2^2\cos^2\theta-4T_1(T_2-T_1)\geq 0. 
\end{equation}
If the TLE experiences no pure dephasing, i.e., $T_2=2T_1$, we can get $\cos^2\theta\geq 1$. Therefore, $\theta=0$ or $\theta=\pi$. Based on the inequality \xj{\eqref{ineq1}}, we can deduce $\theta=0$ and  $x=2\Omega^2 T_1^2$. 
In this scenario, 
the density matrix of the joint TLE-photon system becomes
\begin{equation}
		\hat{\rho}_{ap}=\frac{1}{(1+2\Omega^{2}T_{1}^{2})}\left(\begin{array}{lll}
			1& \Omega T_{1} & \Omega T_{1} \\
			\Omega T_{1} & \Omega^{2} T_{1}^{2} & \Omega^{2} T_{1}^{2} \\
			\Omega T_{1} & \Omega^{2} T_{1}^{2} & \Omega^{2} T_{1}^{2}
		\end{array}\right),
\end{equation}
and its three eigenvalues are $0$, $0$ and $1$, respectively. 
Therefore, the joint state of the two-level emitter and and its spontaneous emission can be written in the pure-state form as 
\begin{equation}
\begin{aligned}\label{}
\ket{\psi}=\sqrt{p_{0}}\ket{g}\ket{0}+\sqrt{p_{1}}\frac{\ket{e}\ket{0}+\ket{g}\ket{1}}{\sqrt{2}},
\end{aligned}
\end{equation}
where 
\begin{equation}
\begin{aligned}\label{}
p_{0}&=\frac{1}{1+2\Omega^{2}T_{1}^{2}},\\
p_{1}&=\frac{2\Omega^{2}T_{1}^{2}}{1+2\Omega^{2}T_{1}^{2}}.
\end{aligned}
\end{equation}
Using this solution, we can see that $p_1$ is equivalent to
	\begin{equation}
		p_1= \frac{2 \bar{n} \eta_{a b}} {1+2 \bar{n} \eta_{a b}},
	\end{equation}
	%as derived through steady-state equilibrium in Main Text,
	where $\bar{n} \eta_{a b}=\Omega^2 T_{1}^{2}$ is the effective average number of photons exciting the TLE while $\bar{n}$ and $\eta_{ab}$ are defined in Main Text as the average incident pump flux and the absorption efficiency under low pump limit, respectively.

\hly{We now transition from the interaction picture to the Schrödinger picture, the pure state should be rewritten as
\begin{equation}
\begin{aligned}\label{}
\ket{\psi}_{ap}=\sqrt{p_{0}}\ket{g}\ket{0}+\sqrt{p_{1}}e^{-\textbf{i}2\pi\nu t}\frac{\ket{e}\ket{0}+\ket{g}\ket{1}}{\sqrt{2}},
\end{aligned}
\end{equation}
where $\nu$ is the frequency of the driving field.}

Finally, we remark that in the case of classical driving field and no pure dephasing, the joint quantum state of the two-level emitter and the spontaneously emitted photon, being a pure state, is intuitive. This is because spontaneous emission merely transfers the coherence of the two-level system to the photon-number state as shown previously in pulsed experiments~\cite{loredo2019generation,wein2022photon}.
However, caution must be had when describing measured quantities using this
pure-state model because it is only valid so long as the TLE system remains in the steady state. That is, any delay lines must be much longer than the relaxation time of the system but still within the coherence time of the driving laser.}

\section {Channel loss}

In any resonance fluorescence (RF) setup, there is inevitably channel loss from the source to the final detection. Here we prove that the channel loss does not affect the analysis of the first or second-order correlation functions, irrespective whether the light field is described by a pure or a mixed state. 

Consider a quantum channel with transmission efficiency $\eta$, the corresponding light field operator transformation relation is 
\begin{equation}
x_{t}\rightarrow\sqrt{\eta}y_{t}+\sqrt{1-\eta}z_{t},
\end{equation}
where $x_{t}$ is annihilation operator
of the input light field, $y_{t}$ and $z_{t}$ are the annihilation operator of transmission and reflection part in quantum channel, respectively.
Tracing over mode $z_{t}$ leads to:
\begin{equation}
\begin{aligned}
\langle x^{\dag}_{t'}x_{t}\rangle&\rightarrow\eta\langle y^{\dag}_{t'}y_{t}\rangle, \\
\langle x^{\dag}_{t}x_{t}\rangle&\rightarrow\eta\langle y^{\dag}_{t}y_{t}\rangle,\\
\langle x^{\dag}_{t}x^{\dag}_{t'}x_{t'}x_{t}\rangle&\rightarrow\eta^{2}\langle y^{\dag}_{t}y^{\dag}_{t'}y_{t'}y_{t}\rangle. 
 \end{aligned}
 \end{equation}
Then, we can calculate the first-order correlation function
\begin{equation}
\begin{aligned}
{\color{black}g^{(1)}(\tau)=\frac{\langle x^{\dag}_{t}x_{t+\tau}\rangle}{\langle x^{\dag}_{t}x_{t}\rangle}=\frac{\langle y^{\dag}_{t}y_{t+\tau}\rangle}{\langle y^{\dag}_{t}y_{t}\rangle}},
 \label{g1:loss}
 \end{aligned}
 \end{equation}
and the second-order correlation function 
\begin{equation}
%\begin{aligned}
g^{(2)}(\Delta t)=\frac{\langle x^{\dag}_{t} x^{\dag}_{t+\Delta t}x_{t+\Delta t}x_{t}\rangle}{\langle x^{\dag}_{t}x_{t}\rangle^{2}}
=\frac{\langle y^{\dag}_{t}y^{\dag}_{t+\Delta t}y_{t+ \Delta t}y_{t}\rangle}{\langle y^{\dag}_{t}y_{t}\rangle^{2}}.\
 \label{g2:loss}
% \end{aligned}
 \end{equation}
Equations~(\ref{g1:loss}, \ref{g2:loss}) mean that channel loss does not affect the analysis of the first- or second-order correlation functions. 
Therefore, we will ignore all channel losses in our subsequent derivation of the RF's coherence.

%%%%%%%%%%%%%%%%%%%%%%
%%%%%%%%%%%%%%%%%%%%%%%

\section {The first-order correlation function $g^{(1)}(\tau)$ and optical frequency spectrum of resonance fluorescence}

\begin{figure*}%[htbp]
\centering
\includegraphics[width=.82\textwidth]{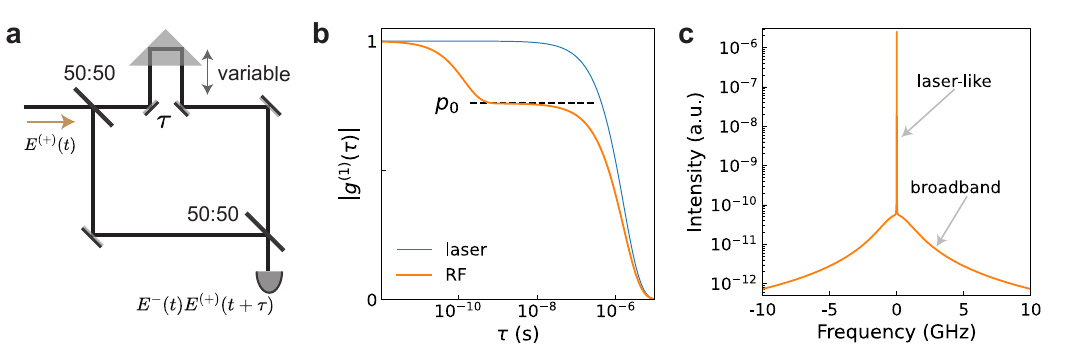}
	\caption{\textbf{The first-order coherence function and optical frequency spectrum.} \textbf{a,} An experimental schematic diagram that can be used to measure first-order correlation function $g^{(1)}(\tau)$. \textbf{b}, Theoretical $g^{(1)}(\tau)$ of RF under a finite excitation power. The curve reveals that the double-exponential decay of $|g^{(1)}(\tau)|$ $=p_1\exp(-\tau/T_2)+p_0\exp(-\tau/T_L)$. As a comparison, we also provide the theoretical $|g^{(1)}(\tau)|=\exp(-\tau/T_L)$ of the laser. \textbf{c}, RF spectrum obtained by performing Fourier transform on the first-order correlation function $|g^{(1)}(\tau)|$ shown in panel \textbf{b}. The laser-like component of the spectrum corresponds to the slowly decaying portion in $|g^{(1)}(\tau)|$, while the broadband corresponds to the rapidly decaying portion of $|g^{(1)}(\tau)|$. In the calculation, we use $T_2 = 137.4$~ps and $T_L = 1.59$~$\mu$s, corresponding to our experimental parameters of QD's lifetime of 67.2~ps and the laser's linewidth of 100~kHz.}
	\label{fig:g1}
\end{figure*}

The first-order correlation function $g^{(1)}(\tau)$  characterises the coherence of an optical field~\cite{Steck2023}.   It measures the normalised interference outcome of an optical field with its delayed copy by time $\tau$, 
\begin{equation}
    {\color{black}g^{(1)}(\tau) = \frac{\langle \hat{E}^{(-)}_{t} \hat{E}^{(+)}_{t+\tau}\rangle}{\langle \hat{E}^{(-)}_{t} \hat{E}^{(+)}_{t}\rangle}},
    \label{eq_SI:g1}
\end{equation}
\noindent where $\hat{E}^{(+)}_{t} = E_0 a'_{t}$ and $\hat{E}^{(-)}_{t} = E_0 a'^\dagger_{t}$ with $a'_{t}$ and $a'^\dagger _{t}$ being photon annihilation and creation operators and $E_0$ the electric field per photon.
A light source is said to be incoherent if $|g^{(1)}(\tau)| = 0$ and coherent if $|g^{(1)}(\tau)| = 1$ for $\tau \neq 0$.
For example, a monochromatic light is perfectly coherent as  $|g^{(1)}(\tau)| \equiv 1$. 
With knowledge of its $g^{(1)}(\tau)$, the frequency spectrum of a light source can be calculated using Wiener-Khinchin theorem,
\begin{equation}
    I(\nu) = \int_0^\infty g^{(1)}(\tau)e^{\textbf{i}2\pi\nu \tau} d \tau, 
\end{equation}
where $\nu$ denotes optical frequency.  A coherent source has a $\delta$-function like spectrum. 
Experimentally, $g^{(1)}(\tau)$ can be measured through an asymmetric Mach-Zehnder interferometer (AMZI) with a variable delay $\tau$, as schematically shown in Supplementary Fig.~\ref{fig:g1}a. Its absolute value represents the interference visibility: $V \equiv |g^{(1)}(\tau)|$.

We now derive the first-order correlation function for the RF of a quantum two-level emitter (TLE) driven by a continuous-wave laser with a coherence time of $T_L$. 
Here, we start with an ideal scenario, i.e., the emitter is free from any extrinsic dephasing process ($T_2 = 2T_1$), and the RF signal is free from any laser background. As derived earlier using master equation model, we write the entangled state between the RF and the TLE under steady-state condition as
\begin{equation} 
\begin{aligned}
       \ket{\psi}_{t}
       &= \sqrt{p_{0}}\ket{0}_{t}\ket{g}_{t}+ \sqrt{p_{1}} {\color{black}e ^ {-\textbf{i}2\pi\nu t}} \frac{\ket{0}_{t}\ket{e}_{t}+\ket{1}_{t}\ket{g}_{t}}{\sqrt{2}},
    \label{eq}
\end{aligned}
\end{equation}
where $\ket{0}_{t}$ denotes vacuum state while $\ket{1}_{t}$ represents a single photon  emitted into temporal mode $t$. $p_0 + p_1 = 1$ and with optical excitation we have $p_0 < 1$.

\zy{For $T_1 \ll \tau \ll T_L$, the state at $t$ and  $t +\tau$ can be considered independent, we can plug} the above TLE-photon state into equations~(\ref{g1:loss})) and  obtain the first-order correlation function, 
\begin{equation}
\begin{aligned}
{\color{black} g^{(1)}(\tau)  =\frac{\bra{\psi}_{t+\tau}\bra{\psi}_{t}a'^{\dag}_{t}a'_{t+\tau}\ket{\psi}_{t}\ket{\psi}_{t+\tau}}{\bra{\psi}_{t}a'^{\dag}_{t}a'_{t}\ket{\psi}_{t}}=p_0 e^{-\textbf{i}2\pi\nu \tau}}.
\label{g11}
\end{aligned}
\end{equation}
\zy{For the second-order correlation function, we directly use the TLE-photon state at time $t$ and obtain}
\begin{equation}
g^{(2)}(0) =\frac{\bra{\psi}_{t}a'^{\dag}_{t}a'^{\dag}_{t}a'_{t}a'_{t}\ket{\psi}_{t}}{\bra{\psi}_{t}a'^{\dag}_{t}a'_{t}\ket{\psi}_{t}^{2}}=0,
\label{g2}
\end{equation}
where $a'^{\dag}_{t}$ and $a'^{\dag}_{t+\tau}$ act in quantum states $\ket{\psi}_{t}$ and $\ket{\psi}_{t+\tau}$, respectively. The RF has a finite coherence ($p_0 < 1$) and simultaneously photon anti-bunching, which is in agreement with experiments.

For short delays, i.e., $\tau \in [0, T_2]$, the first-order
correlation function measures the interference of
each RF photon with itself. Therefore, $g^{(1)}$ is governed by the TLE’s dephasing time
$T_2$ under weak excitation, and will become
mediated by the Rabi oscillation under strong excitation. Then, we have $|g^{(1)}(0)| = 1$ at $\tau = 0$, followed by a fast decay to the plateaued value of $p_0$.  When $\tau$ becomes comparable to or exceeds the laser coherence time $T_L$, $|g^{(1)}(\tau)|$ will start its second exponential decay.  The overall dependence on $\tau$ is illustrated in Fig.~\ref{fig:g1}\textbf{b}. 

Through Fourier transform, we can calculate the RF frequency spectrum as shown in Supplementary Fig.~\ref{fig:g1}\textbf{c}. Closely resembling the experimental data (Fig.~1\textbf{d}, Main Text), the spectrum contains a sharp peak that inherits the laser linewidth and a broadband pedestal of the TLE’s bandwidth. The laser-like ($ll$) part has a spectral weight of $I_{ll}/I_{tot} = |g^{(1)}(\tau)| =  p_0 < 1$.

\section{Origin of the RF's coherence and its partial loss}

To reveal how the RF obtains or loses its coherence, we derive the interference outcome when passing it through an AMZI ($T_1 \ll \tau \ll T_L$).  As two-path interference occurs between a late temporal mode $\ket{\psi}_t$ passing through the short arm and an early mode $\ket{\psi}_{t-\tau}$ through the long arm,  we derive from equation~(\ref{eq}) the joint input state between two temporal modes as a tensor product
%\begin{widetext}
\begin{equation}
\begin{aligned}
\ket{\psi}_{t-\tau}\ket{\psi}_{t}&=\ket{00}\left[p_{0}\ket{gg}+\frac{\sqrt{p_{0}p_{1}}}{\sqrt{2}}(\ket{ge}+\ket{eg}) +\frac{p_{1}}{2}\ket{ee}\right]\\
&+\frac{\sqrt{p_{0}p_{1}}}{\sqrt{2}}\left(\ket{10}+\ket{01}\right)\ket{gg}\\
&+\frac{p_{1}}{2}(\ket{10}\ket{ge}+\ket{01}\ket{eg}) +\frac{p_{1}}{2}\ket{11}\ket{gg},
 \label{Eq_SI:tensor_state}
 \end{aligned}
 \end{equation}
%\end{widetext}
where all time evolution phases are dropped for clarity. State $\ket{01}$ represents 0-photon at temporal mode $t-\tau$ and 1-photon at mode $t$, state $\ket{ge}$ denotes the TLE occupying the ground state at time $t-\tau$ and the excited state at time $t$. All other \xj{ket} states are defined in the same way.

The first term in equation~(\ref{Eq_SI:tensor_state}) contains no photons. The second term contains $\ket{10}+\ket{01}$, which represents a typical single-photon entanglement between two temporal modes and will produce interference between $\ket{10}$ passing through the long arm and $\ket{01}$ the short arm. This interference is the origin of the RF's coherence.
 
To derive the AMZI output, we use port labelling \zy{($a'$, $a$, $b$, $c$ and $d$)}, as shown in Fig.~1\textbf{b} of Main Text \zy{and also later in Supplementary Fig.~\ref{setup2}.} 
The transformation relation between input and output light field annihilation operators of the AMZI can be given by 
\begin{equation}
\begin{aligned}
a'_{t}&=\frac{1}{2}\left\{[d_{t+\tau}+c_{t+\tau}]+e^{-\textbf{i}\varphi}[d_{t}-c_{t}]\right\},\\
a'_{t-\tau}&=\frac{1}{2}\left\{[d_{t}+c_{t}]+e^{-\textbf{i}\varphi}[d_{t-\tau}-c_{t-\tau}]\right\}.\\
\label{}
\end{aligned}
\end{equation} 
The  joint output quantum state of ports $c$ and $d$ at time $t$ is the interference result of the RF state at times $t-\tau$ and $t$, see equation~(\ref{Eq_SI:tensor_state}). 
Note that we have proved in Supplementary Section~I that the attenuation along the optical path does not affect the RF's first-order coherence. Therefore, for clarity of expression, we treat the first beam splitter of the AMZI as a 3~dB attenuation without changing the form of the quantum state\cite{loredo2019generation}. 
For simple calculation and intuitive understanding, we use pure state analysis without considering channel loss and the first beam splitter of the AMZI. 
This treatment is equivalent to interfering the RF field with a delayed copy of itself, thus simplifying the derivation of the AMZI's output state.
The simplified operator transformation relation is as follows (normalized)
\begin{equation}
\begin{aligned}
a'_{t}&=\frac{1}{\sqrt{2}}e^{-\textbf{i}\varphi}[d_{t}-c_{t}],\\
a'_{t-\tau}&=\frac{1}{\sqrt{2}}[d_{t}+c_{t}].\\
\label{}
\end{aligned}
\end{equation}
With the input state described in equation~\eqref{Eq_SI:tensor_state}, the normalized output joint quantum state of ports $c$ and $d$ at time $t$ then becomes 
\begin{widetext}
\begin{equation}
\begin{aligned}%\label{intfer}
\ket{\Psi_{\rm out} }_{t}= &\ket{0_c0_d}_t\left(p_{0}\ket{gg}+\frac{\sqrt{p_{0}p_{1}}}{\sqrt{2}}\ket{ge}+\frac{\sqrt{p_{0}p_{1}}}{\sqrt{2}}\ket{eg}+\frac{p_{1}}{2}\ket{ee}\right)\\
+&\ket{1_c0_d}_t\frac{\sqrt{p_{0}p_{1}}}{\sqrt{2}}\frac{1-e^{\textbf{i}\varphi}}{\sqrt{2}}\ket{gg}
+\ket{1_c0_d}_{t}\frac{p_{1}}{2\sqrt{2}}\left(\ket{ge}-e^{\textbf{i}\varphi}\ket{eg} \right) -\ket{2_c0_d}_t e^{\textbf{i}\varphi}\frac{p_{1}}{2\sqrt{2}}\ket{gg}\\
+&\ket{0_c1_d}_t\frac{\sqrt{p_{0}p_{1}}}{\sqrt{2}}\frac{1+e^{\textbf{i}\varphi}}{\sqrt{2}}\ket{gg}+\ket{0_c1_d}_{t}\frac{p_{1}}{2\sqrt{2}}\left(\ket{ge}+e^{\textbf{i}\varphi}\ket{eg} \right)  +\ket{0_c2_d}_t e^{\textbf{i}\varphi}\frac{p_{1}}{2\sqrt{2}}\ket{gg},\\
\label{eq:output}
\end{aligned}
\end{equation}
\end{widetext}
when all RF photons are indistinguishable.
%Equation~(\ref{eq:output}) tells us how the RF gains and loses its coherence. 
The first line contains no photons, while the second and third lines represent \xj{complementary} outputs at ports $c$ and $d$.  Let's look just at port $c$. It contains a phase-dependent term $\ket{1_c}_t\ket{gg}$, corresponding to the TLE's transition from $(\ket{ge} + \ket{eg})/\sqrt{2} \to \ket{gg}$. Varying the AMZI's phase $\varphi$, this term produces interference fringes with an amplitude of $\frac{p_0p_1}{2}$.  
The remaining terms are non-interfering because $\ket{1_c}\ket{ge}$, $\ket{1_c}\ket{eg}$ and $\ket{2_c}\ket{gg}$ are projected onto different matter states.  The corresponding transitions are from the same $\ket{ee}$ state, to different final states $\ket{ge}$, $\ket{eg}$ and $\ket{gg}$ respectively.  The first two transitions correspond to the TLE emitting one photon into just one of the two temporal modes, while the last one means emitting one photon to each temporal mode and Hong-Ou-Mandel (HOM) interference produces the two-photon term $\ket{2_c}$.
Altogether, these non-interfering terms contribute $\frac{p_1^2}{8} \times 2 + \frac{p_1^2}{4} = \frac{p_1^2}{2}$. Therefore, we obtain an interference fringe visibility as 
$\frac{p_0p_1}{2}/(\frac{p_0p_1}{2} + \frac{p_1^2}{2}) =  p_0$, which is identical to the earlier result of equation~(\ref{g11}).  To sum up, the RF gains and loses its coherence both through spontaneous emission.

The discussion above is based on the assumption that single photons emitted by the TLE are perfectly indistinguishable. However, photons scattered out by the TLE underwent spontaneous emission processes, and therefore their indistinguishability would be degraded by the TLE's extrinsic scattering processes that are inherent in solid-state quantum systems.
Taking into account of photon distinguishability, the first order coherence is revised  
accordingly to~\cite{loredo2019generation} 
\begin{equation}
\left|g^{(1)}_{RF}(\tau)\right|=\sqrt{M}p_0,
\label{g1}
\end{equation}
where $M$ represents the indistinguishability of single photons between different temporal modes, 
\begin{equation}
    M=\left|\langle 1_t|1_{(t-\tau)\rightarrow t}\rangle \right|^2,
    \label{M:defination}
\end{equation}
 where $(t-\tau)\rightarrow t$ means 
 delaying  mode $t - \tau$ by $\tau$ for temporally overlapping $\ket{1_t}$ and $\ket{1_{t-\tau}}$. 

\begin{figure}[h]
\includegraphics[width=.7\columnwidth]{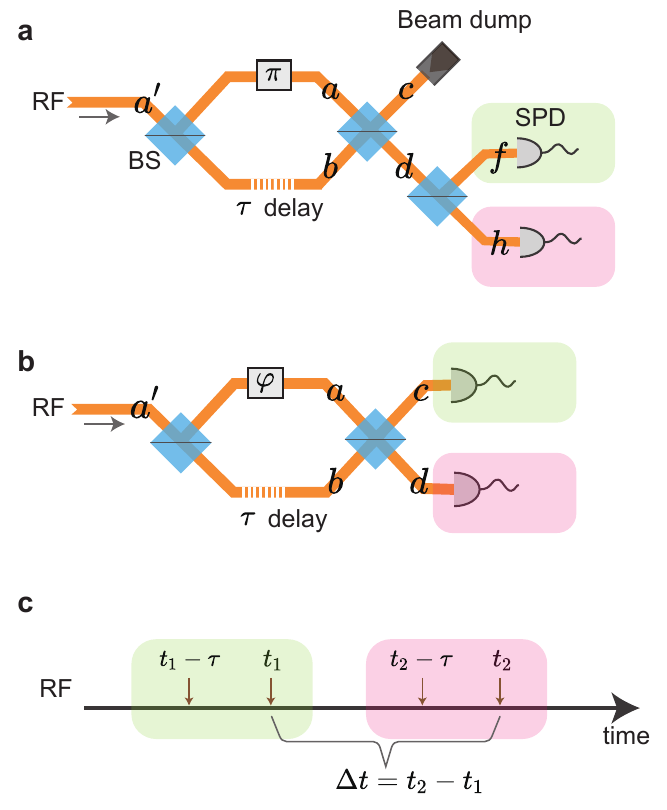}
	\caption{ \textbf{Phase-dependent second-order correlations.} \textbf{a} $g^{(2)}(\Delta t)$ measurement setup for RF after $\pi$-phase AMZI filtering. \textbf{b}, HOM interferometer with a variable phase delay $\varphi$.  \textbf{c}, Due to the AMZI delay $\tau$, each coincidence count is related to up to four emission times at the RF source.} 
	\label{setup2}
\end{figure}

\section{Photon super-bunching in the AMZI-filtered RF}

According to equation~\eqref{eq:output}, all the interfering  single-photon fraction (laser-like) will exit from the port $c$ if we set the AMZI's phase to $\varphi = \pi$.  With the laser-like spectrum rejected, the port $d$ will then contain two non-interfering single photon terms $\ket{1_d}\ket{ge}$ and $\ket{1_d}\ket{eg}$ as well as a two-photon term $\ket{2_d}\ket{gg}$. 
Measured with an auto-correlation setup as shown in Supplementary Fig.~2\textbf{a}, the port $d$ output will exhibit super-bunching that arises from the two-photon term.  Note this super-bunching is simply a result of two-photon  interference and does not necessitate simultaneous scattering of two photons~\cite{masters2023} for an explanation.

In the HBT setup (Supplementary Fig.~2\textbf{a}), a coincidence event with $f$ detector clicked at $t_1$ and $h$ at $t_2$ involves the input of up to four non-degenerate temporal modes, $t_1$, $t_1-\tau$, $t_2$ and $t_2-\tau$, see Supplementary Fig.~2\textbf{c}. Below, we will derive the second-order correlation according to the level of their time degeneracy.

\subsection{Non-degenerate}
For interval $\Delta t=t_2-t_1$ that meets both $|\Delta t \pm \tau| \gg T_1$ and $|\Delta t| \gg T_1$, there is no degeneracy in the four involved temporal modes. In Schrodinger picture, the second-order correlation function at $\Delta t\neq 0,\pm\tau $ is given by
\begin{equation}
\begin{aligned}
&g_{\varphi =\pi}^{(2)}(\Delta t)\\
&=\frac{\bra{\Psi_{\rm out}}_{t_{2}}\bra{\Psi_{\rm out}}_{t_{1}}d^{\dag}_{t_{2}}d^{\dag}_{t_{1}}d_{t_{1}}d_{t_{2}}\ket{\Psi_{\rm out}}_{t_{1}}\ket{\Psi_{\rm out}}_{t_{2}}}{\bra{\Psi_{\rm out}}_{t_{1}}d^{\dag}_{t_{1}}d_{t_{1}}\ket{\Psi_{\rm out}}_{t_{1}}^{2}}\\
&=\frac{\bra{\Psi_{\rm out}}_{t_{2}}d^{\dag}_{t_{2}}d_{t_{2}}\ket{\Psi_{\rm out}}_{t_{2}}\bra{\Psi_{\rm out}}_{t_{1}}d^{\dag}_{t_{1}}d_{t_{1}}\ket{\Psi_{\rm out}}_{t_{1}}}{\bra{\Psi_{\rm out}}_{t_{1}}d^{\dag}_{t_{1}}d_{t_{1}}\ket{\Psi_{\rm out}}_{t_{1}}^{2}}\\
&=1,
\end{aligned}
\end{equation}
where state $\ket{\Psi_{\rm out}}_{t}$ is given by in equation \eqref{eq:output} and $\ket{\Psi_{\rm out}}_{t_{1}}\ket{\Psi_{\rm out}}_{t_{2}}$ is the  tensor product state between $t_{1}$ and $t_{2}$.

\subsection{Doubly degenerate: $t_1 = t_2$}

At $\Delta t = 0$, the four temporal modes become doubly degenerate: $t_1 = t_2=t$ and $t_1-\tau= t_2-\tau=t-\tau$. we work out the second-order correlation function for the 0-delay:
\begin{equation}
g_{\varphi = \pi}^{(2)}(0)  =\frac{\bra{\Psi_{\rm out}}[d^{\dag}_{t}]^{2}d^{2}_{t}\ket{\Psi_{\rm out}}}{\bra{\Psi_{\rm out}}d^{\dag}_{t}d_{t}\ket{\Psi_{\rm out}}^{2}}=\frac{1}{p_{1}^{2}}.
\label{eq_SI:HBT_AMZI}
\end{equation}
We obtain super-bunching at 0-delay and it is excitation power dependent.  
The lower the excitation power, the higher the $g_{\varphi = \pi}^{(2)}(0)$ value becomes.
At the limit of $\bar{n} \to 0$, $g_{\varphi = \pi}^{2}(0)$ can be infinitely large in theory. We emphasise again that this super-bunching does not require a higher-order scattering mechanism, contrary to previously suggested~\cite{masters2023}.

\subsection{Singly-degenerate: $t_2 - t_1 = \pm\tau$}

For time intervals $t_{2}-t_{1}=\Delta t = \pm \tau$, three temporal modes at the AMZI input contribute to the coincidence at $\Delta t = \pm \tau$.  
For $\Delta t = \tau$, the corresponding modes are  $t-\tau$, $t$ and $t+\tau$.
Input temporal mode $t$ is split into two halves by the AMZI entrance beam splitter. The half passing through the short (long) arm interferes half of temporal mode $t-\tau$ ($t + \tau$) passing through the long (short) arm. At the AMZI output $d$, we just need to consider two temporal modes $t$ and $t + \tau$ for calculating its $g^{(2)}(+\tau)$.
We use the transformation relation (normalized) between output and input light field annihilation operators of the AMZI with $\varphi=\pi$% can be rewritten as (normalized)
\begin{equation}
\begin{aligned}
d_{t}&=\frac{1}{\sqrt{2}}\left[a'_{t-\tau}-a'_{t}\right],\\
d_{t+\tau}&=\frac{1}{\sqrt{2}}\left[a'_{t}-a'_{t+\tau}\right],\\
\label{18}
\end{aligned}
\end{equation}
\xj{where we ignore the loss by the first beam splitter, and $a^\prime$ denotes the input port of the AMZI, as labeled in Fig.~1b in Main Text and Supplementary Fig.~2}.
Using the Heisenberg picture, the second order correlation function can be derived using 
\begin{equation}
\begin{aligned}
g_{\varphi =\pi}^{(2)}(\tau)&=\frac{\bra{\psi_{\rm in}}d^{\dag}_{t+\tau}d^{\dag}_{t}d_{t}d_{t+\tau}\ket{\psi_{\rm in}}}{\bra{\psi_{\rm in}}d^{\dag}_{t}d_{t}\ket{\psi_{\rm in}}^{2}}.
\end{aligned}
\end{equation}
where operators evolve according to equation~\eqref{18} and we have $\ket{\psi_{\rm in}}=\ket{\psi}_{t-\tau}\ket{\psi}_{t}\ket{\psi}_{t+\tau}$.
With time-evolving phases dropped for clarity, we have
\begin{widetext}
\begin{equation}
\begin{aligned}
d_{t}\ket{\psi_{\rm in}}&=\frac{1}{\sqrt{2}}\left[a'_{t-\tau}-a'_{t}\right]\ket{\psi}_{t-\tau}\ket{\psi}_{t}\ket{\psi}_{t+\tau}=\frac{p_{1}}{2\sqrt{2}}[\ket{00}(\ket{ge}+\ket{eg})+(\ket{01}+\ket{10})\ket{gg}]\ket{\psi}_{t+\tau}.
\label{}
\end{aligned}
\end{equation}
and
\begin{equation}
\begin{aligned}
d_{t}d_{t+\tau}\ket{\psi_{\rm in}}&=\frac{1}{2}\left[a'_{t-\tau}-a'_{t}\right]\left[a'_{t}-a'_{t+\tau}\right]\ket{\psi}_{t-\tau}\ket{\psi}_{t}\ket{\psi}_{t+\tau}\\
&=\frac{p_{1}}{4}\left\{\sqrt{p_{0}}\ket{000}\ket{ggg}+\frac{\sqrt{p_{1}}}{\sqrt{2}}\big[\ket{000}(\ket{gge}-\ket{geg}+\ket{gge})+(\ket{001}-\ket{010}+\ket{100})\ket{ggg}\big]\right\}.
\label{}
\end{aligned}
\end{equation}
\end{widetext}
We then obtain the second order correlation function at $\Delta t=\tau$:
\begin{equation}
\begin{aligned}
g_{\varphi=\pi}^{(2)}(\tau)=\frac{p_{1}^{2}(1+2p_{1})/16}{(p_{1}^{2}/2)^{2}}=\frac{1+2p_{1}}{4p_{1}^{2}}.
\end{aligned}
\end{equation}

Similarly, we derive the same conclusion for $\Delta t=-\tau$, i.e., $g_{\varphi=\pi}^{(2)}(\tau)=g_{\varphi=\pi}^{(2)}(-\tau)$.
When $p_{1}\ll1$, we can have $1+2p_{1}\approx1$ and the second order interference coherence can be written as
\begin{equation}
\begin{aligned}
g_{\varphi=\pi}^{(2)}(\pm\tau)%&=\frac{1+2p_{1}}{4p_{1}^{2}}\\
%&
\approx\frac{1}{4p_{1}^{2}}=\frac{1}{4}g_{\varphi=\pi}^{(2)}(0).
\end{aligned}
\end{equation}

\section{Phase-dependent Hong-Ou-Mandel interference}

Hong-Ou-Mandel interferometry is an indispensable tool for evaluating indistinguishability among photons emitted by a pulsed single photon source~\cite{Santori2002}. 
As shown in Supplementary Fig.~2\textbf{b}, a typical HOM interferometer consists of an AMZI with a path difference of several nanoseconds and two single photon detectors.  
The AMZI's differential delay ($\tau$) can bring two single photons separated by $\tau$ to temporally overlap for interference. 
If perfectly indistinguishable, two photons will coalesce and therefore the possibility for registering a photon simultaneously at each detector port is 0.

Continuous-wave HOM interference differs from the aforementioned results for pulsed single photons. The reason lies in the fact that when the detector's time resolution is arbitrarily high, unity indistinguishability would be obtained, i.e., $M=1$, even if the two photons' spectral and temporal envelopes are not identical. More interestingly, in our experiments, due to the RF's long first-order coherence, coincidences of $\Delta t=\pm \tau$ are significantly different from those sources having short coherence times
~\cite{Proux2015,PhysRevLett.100.207405}. 
As shown in Fig.~4 of the Main Text, we observe strong dependencies of the coincidence on the AMZI phase and excitation strength.  
Below, we give theoretical derivations of the coincidence $\mathcal{C}(\Delta t)$ for phase- and excitation power dependent HOM interference.

\subsection {General description}

As shown in Supplementary Fig.~\ref{setup2}b, a quantum light input, now described by a density matrix $\rho_{t}$ for convenience, enters the AMZI through port $a'$ and is split equally into two paths $a$ and $b$.  
With path $a$ accumulating a phase of $\varphi$ and path $b$ a delay of $\tau$, the split signals recombine at the exit 50/50 beam splitter to interfere.  Photons exiting from ports $c$ and $d$ are detected by two single photon detectors. The transformation relation between output and input light field annihilation operators of the AMZI can be given by 
\begin{equation}
\begin{aligned}
c_{t}&=\frac{1}{2}\left[a'_{t-\tau}-e^{\textbf{i}\varphi}a'_{t}\right],\\
d_{t}&=\frac{1}{2}\left[a'_{t-\tau}+e^{\textbf{i}\varphi}a'_{t}\right],\\
\end{aligned}
\end{equation}
where we have included the loss by the first beam splitter.

If we drop the phase \xj{$e^{-\textbf{i}2\pi\nu t}$} in equation~\eqref{eq} that does not affect the result of the calculation, the density matrix for describing the RF state can be written as
\begin{equation}
\begin{aligned}
\frac{2p_{0}+p_{1}}{2}\ket{0}\bra{0}+\frac{p_{1}}{2}\ket{1}\bra{1}+\sqrt{\frac{p_{0}p_{1}}{2}}\left(\ket{1}\bra{0}+\ket{0}\bra{1}\right),
\label{}
\end{aligned}
\end{equation}
after performing partial trace over system TLE. The resulting density matrix is equivalent to a pure state $\ket{\phi}=\sqrt{p_{0}}\ket{0}+\sqrt{p_{1}}\ket{1}$ after 3 dB channel loss and performing partial trace over the lost %reflected 
part. As we have proven in Section~I that the channel loss does not affect the first and second order interference coherence,  we can conveniently use the pure state $\ket{\phi}=\sqrt{p_{0}}\ket{0}+\sqrt{p_{1}}\ket{1}$ to calculate the coincidence probabilities in the HOM interference without loss of generality. 

In the following derivation, we will use the pure state 
\begin{equation}
\begin{aligned}
\ket{\phi_t}=\sqrt{p_{0}}\ket{0_t}+\sqrt{p_{1}}\ket{1_t}+\sqrt{p_{2}}\ket{2_t}
\label{}
\end{aligned}
\end{equation}
with $p_0 + p_1 + p_2 = 1$ and $p_2 \ll p_1^2/2$,  instead of a mixture state density matrix as the incident light source for calculation.
The introduction of the two-photon term $\ket{2_t}$ is to reflect the experimental setup imperfection that mixes a small amount of laser photons into the RF.
The two-photon term is significant only for the calculation of the HOM dip. 

\subsection{Coincidence probability}

A coincidence event with $c$ detector clicked at $t_1$ and $d$ at $t_2$ involves the input of up to four non-degenerate times, $t_1$, $t_1-\tau$, $t_2$ and $t_2-\tau$, see Supplementary Fig.~\ref{setup2}c. Below, we will analyse their coincidence probabilities according to the level of their time degeneracy.

\subsubsection{Non-degenerate}

For intervals $\Delta t=t_2-t_1$that meets both $|\Delta t \pm \tau| \gg T_1$ and $|\Delta t| \gg T_1$, there is no degeneracy in the four involved times and each coincidence therefore is a result of  two independent first-order interference  events.
For the incident quantum state, since the contribution of multi-photon components is small and has little impact on the final count, we consider only the lowest-order photon state $|1\rangle$ to capture the main characteristics of the baseline coincidence probability.
The count probabilities of detectors in ports $c$ and $d$ can be written as 
\begin{equation}
\begin{aligned}
\mathcal{P}_{c}&=\bra{\phi_{t_{1}}}\bra{\phi_{t_{1}-\tau}}c^{\dag}_{t_{1}}c_{t_{1}}\ket{\phi_{t_{1}-\tau}}\ket{\phi_{t_{1}}}\\
&=\left|c_{t_{1}}\ket{\phi_{t_{1}-\tau}}\ket{\phi_{t_{1}}}\right|^{2}\\
&=\left|\frac{1}{2}\left[\sqrt{p_{0}p_{1}}\left(1-e^{\textbf{i}\varphi}\right)\ket{00}+p_{1}\left(\ket{01}-e^{\textbf{i}\varphi}\ket{10}\right)\right]\right|^{2}\\
&=\frac{p_{1}}{2}(1-p_{0}\cos\varphi),
\label{}
\end{aligned}
\end{equation}
and 
\begin{equation}
\begin{aligned}
\mathcal{P}_{d}&=\bra{\phi_{t_{2}}}\bra{\phi_{t_{2}-\tau}}d^{\dag}_{t_{2}}d_{t_{2}}\ket{\phi_{t_{2}-\tau}}\ket{\phi_{t_{2}}}\\
&=\frac{p_{1}}{2}(1+p_{0}\cos\varphi),
\label{}
\end{aligned}
\end{equation}
where $a'_{t_{1}}$ and $a'_{t_{1}-\tau}$ act in quantum states $\ket{\phi_{t_{1}}}$ and $\ket{\phi_{t_{1}-\tau}}$, respectively and state $\ket{\phi_{t}}=\sqrt{p_{0}}\ket{0}+\sqrt{p_{1}}\ket{1}$ after ignoring the two-photon component.
Taking into account of imperfect photon indistinguishability defined in equation~\eqref{M:defination}, we have 
\begin{equation}
\begin{aligned}
\mathcal{P}_{c}&=\frac{p_{1}}{2}(1-\sqrt{M}p_{0}\cos\varphi),\\
\mathcal{P}_{d}&=\frac{p_{1}}{2}(1+\sqrt{M}p_{0}\cos\varphi).\\
\label{}
\end{aligned}
\end{equation}
Consequently, the coincidence probability is simply the product of the count probabilities of individual detectors, 
\begin{equation}
\mathcal{C}_0 =\mathcal{P}_{c}
\mathcal{P}_{d} 
    =\frac{p_1^2}{4}\left(1-Mp_0^2\cos^2\varphi\right).
\label{eq:baseline}
\end{equation}
$\mathcal{C}_0$  the coincidence baseline depends on the AMZI's phase.   At $\varphi = \pi/2$, it reaches its maximum. 

Next we give a different \xj{derivation} method to elaborate the correctness of equation~\eqref{eq:baseline}. It will show the special properties of a continuous-wave RF, i.e.,  approximating the RF's quantum state as a coherent superposition of $|0\rangle$ and $|1\rangle$ is an acceptable treatment as long as the examined time interval is much larger than $T_1$.  Rigorously, its coincidence probability can be derived using~\cite{loredo2019generation,legero2003time},
\begin{equation}
	\begin{aligned}
	\mathcal{C}(\Delta t)&=\bra{\psi_{\rm in}} c^\dagger _{t_{1}} c_{t_{1}}d^\dagger_{t_{2}} d_{t_{2}}\ket{\psi_{\rm in}}\\
&=\bra{\psi_{\rm in}} d^\dagger_{t_{2}}c^\dagger_{t_{1}}  c_{t_{1}}d_{t_{2}}\ket{\psi_{\rm in}},\\
	\end{aligned}
\end{equation}
where $\ket{\psi_{\rm in}}$ is the input quantum state and 
\begin{equation}	
	\begin{aligned}
c_{t_{1}}d_{t_{2}}=\frac{1}{4}\left[a'_{t_{1}-\tau}-e^{\textbf{i}\varphi}a'_{t_{1}}\right]\left[a'_{t_{2}-\tau}+e^{\textbf{i}\varphi}a'_{t_{2}}\right].		
	\end{aligned}
\end{equation}
Let $t_1=t$, then $t_2=t+\Delta t$, which leads to
\begin{widetext}
\begin{equation}	
	\begin{aligned}
c_{t_{1}}d_{t_{2}}=c_{t}d_{t+\Delta t}
=& \frac{1}{4}\left[a'_{t-\tau}a'_{t+\Delta t-\tau}  + a'_{t-\tau} a'_{t+\Delta t}e^{\textbf{i} \varphi}-a'_{t}a'_{t+\Delta t-\tau}  e^{\textbf{i} \varphi}-a'_{t}a'_{t+\Delta t}e^{\textbf{i} 2 \varphi}\right] .
	\end{aligned}
 \label{oper}
\end{equation}
Since there are four temporal modes involved in the above operator expression and the time interval between each other is much larger than $T_1$, the input quantum state can be expressed as a tensor product of the four temporal modes, i.e., $\ket{\psi_{\rm in}}=\ket{\phi_{t-\tau}}\ket{\phi_{t}}\ket{\phi_{t+\Delta t-\tau}}\ket{\phi_{t+\Delta t}}$. Therefore,
\begin{equation}
\begin{aligned}
    &c_{t}d_{t+\Delta t}\ket{\psi_{\rm in}}\\
    =& {\left[p_0 p_1\left(-e^{\textbf{i} 2 \varphi}+e^{\textbf{i}\varphi}-e^{\textbf{i} \varphi}+1\right)|0000\rangle+\sqrt{p_0 p_1^3}\left(1+e^{\textbf{i} \varphi}\right)|0100\rangle\right.} \\
& +\sqrt{p_0 p_1^3}\left(-e^{\textbf{i} \varphi}-e^{\textbf{i} 2 \varphi}\right)|1000\rangle+\sqrt{p_0 p_1^3}\left(1-e^{\textbf{i} \varphi}\right)|0001\rangle \\
& \left. +\sqrt{p_0 p_1^3}\left(e^{\textbf{i} \varphi}-e^{\textbf{i} 2 \varphi}\right)|0010\rangle+p_1^2 |0101\rangle+p_1^2 e^{\textbf{i} \varphi}|0110\rangle -p_1^2 e^{\textbf{i}\varphi}|1001\rangle+p_1^2e^{\textbf{i} 2 \varphi}|1010\rangle\right] / 4,
\end{aligned}
\end{equation}
thereby 
\begin{equation}
\begin{aligned}
\mathcal{C}(\Delta t)=&\bra{\psi_{\rm in}} d^\dagger_{t_{2}}c^\dagger_{t_{1}}  c_{t_{1}}d_{t_{2}}\ket{\psi_{\rm in}} =\frac{1}{4} p_0^2 p_1^2 \sin ^2 \varphi+\frac{1}{2} p_0 p_1^3+\frac{1}{4} p_1^4.
\end{aligned}
\end{equation}
\end{widetext}
Considering the influence of photon indistinguishability, $\mathcal{C}(\Delta t)$ should be corrected as
\begin{equation}
\begin{aligned}
\mathcal{C}(\Delta t)= &\frac{1}{4} p_0^2 p_1^2\left(1-M \cos ^2 \varphi\right)+\frac{1}{2} p_0 p_1^3+\frac{1}{4} p_1^4\\ 
%=& \frac{1}{4}p_1^2 \left[ (p_0+p_1)^2-Mp_0^2\cos^2\varphi\right] \\
=&\frac{1}{4}p_1^2 \left[ 1-Mp_0^2\cos^2\varphi\right]=\mathcal{C}_0.
\end{aligned}
\end{equation}
We obtain the same result as equation~\eqref{eq:baseline}.

\subsubsection{Singly-degenerate: $t_2 - t_1 = \pm\tau$}

For intervals $\Delta t = \pm \tau$,  two out of the four time slots  become degenerate. Let's first consider $\Delta t = +\tau$. In this case,  we write the operator corresponding based on equation (\ref{oper}),
\begin{equation}
c_{t}d_{t+\tau}=\frac{1}{4}\left[a'_{t-\tau}a'_{t}+e^{\textbf{i}\varphi}a'_{t-\tau}a'_{t+\tau}-e^{\textbf{i}\varphi}a'_{t}a'_{t}-e^{\textbf{i}2\varphi}a'_{t}a'_{t+\tau}\right].
\end{equation}
Since the contribution of multi-photon components has negligible contribution to the coincidence at this delay, we use quantum state $\ket{\phi_{t}}=\sqrt{p_{0}}\ket{0}+\sqrt{p_{1}}\ket{1}$ to capture the main characteristics of the coincidence probability.

According to the expression of $c_{t}d_{t+\tau}$, the output state of the AMZI at time $t$ is the interference result between the input state at time $t-\tau$ taking the long path and the input state at time $t$ taking the short path. Accordingly, the output state of the AMZI at time $t+\tau$ is the interference result between the input state at time $t$ taking the long path and the input state at time $t+\tau$ taking the short path. The input state for deriving $\mathcal{C}(+\tau)$, i.e., the coincidence probability at $+\tau$ interval between detectors $c$ and $d$, 
should therefore be the tensor product of quantum states of three times $t-\tau$, $t$ and $t+\tau$, i.e.,
$\ket{\psi_{\rm in}}=\ket{\phi_{t-\tau}}\ket{\phi_{t}}\ket{\phi_{t+\tau}}$
\begin{widetext}
\begin{equation}
	\begin{aligned}
c_{t}d_{t+\tau}\ket{\phi_{t-\tau}}\ket{\phi_{t}}\ket{\phi_{t+\tau}}=\frac{1}{4}\left[p_{1}\sqrt{p_{0}}\left(1+e^{\textbf{i}\varphi}-e^{\textbf{i}2\varphi}\right)\ket{000}+\sqrt{p_{1}^{3}}\left(\ket{001}+e^{\textbf{i}\varphi}\ket{010}-e^{\textbf{i}2\varphi}\ket{100}\right)\right],
	\end{aligned}
\end{equation}
\end{widetext}
where $a'_{t+\tau}$, $a'_{t}$ and $a'_{t-\tau}$ act on states $\ket{\phi_{t-\tau}}$, $\ket{\phi_{t}}$, and $\ket{\phi_{t+\tau}}$, respectively. 
We then obtain the coincidence probability
\begin{equation}
	\mathcal{C}(\tau)=\frac{1}{16}p_0p_1^2(3-2\cos 2\varphi)+\frac{3}{16}p_1^3.
\end{equation}

For $\Delta t = - \tau$, one can follow the same derivation process and obtain $\mathcal{C}(-\tau) = \mathcal{C}(+\tau)$. 
Taking into account of imperfect photon indistinguishability, $\mathcal{C}(\pm\tau)$ is corrected to
\begin{equation}
		\mathcal{C}(\pm\tau)=\frac{1}{16}p_0p_1^2(3-2M\cos 2\varphi)+\frac{3}{16}p_1^3.
\end{equation}

%%%%%%%%%%%%%%%%%%%%
%%%%%%%%%%%%%%%%%%%%

\subsubsection{Doubly degenerate: $t_1 = t_2$}

At $\Delta t = 0$, the four time slots (Supplementary Fig.~2c) become doubly degenerate: $t_1 = t_2=t$ and $t_1-\tau = t_2- \tau=t-\tau$.  Coincidence at this interval has two contributions: (1) multi-photon components in the input and (2) imperfect two-photon interference.  As these two contributions are on the same magnitude, it is necessary to include the two-photon term in the input state in order to derive the correct coincidence probability. 
We use the pure state $\ket{\phi_{t}}=\sqrt{p_{0}}\ket{0}+\sqrt{p_{1}}\ket{1}+\sqrt{p_{2}}\ket{2}$ as the input. 

Following equation (\ref{oper}), the operator for $\Delta t=0$ and $t_{1}=t_{2}=t$ can be written as
\begin{equation}
	\begin{aligned}
c_{t}d_{t}&=\frac{1}{4}\left[a'_{t-\tau}a'_{t-\tau}-e^{\textbf{i}2\varphi}a'_{t}a'_{t}+\underline{e^{\textbf{i}\varphi}a'_{t-\tau}a'_{t}-e^{\textbf{i}\varphi}a'_{t}a'_{t-\tau}}\right],
	\end{aligned}
\end{equation}
where the first two terms produce interference between an early two-photon state taking the long arm and a late two-photon state passing the short arm, while the underlined terms cause two-photon HOM interference between an early and a late single photon.  If photons are not identical and the detector resolution is limited, the output by the underlined terms will not cancel completely. According the above equation, we only need to consider the states of two input temporal modes $t-\tau$ and $t$, i.e., $\ket{\psi_{\rm in}}=\ket{\phi_{t-\tau}}\ket{\phi_{t}}$ leading to
\begin{widetext}
\begin{equation}
\begin{aligned}
c_{t}d_{t}\ket{\phi_{t-\tau}}\ket{\phi_{t}}=&\frac{1}{4}\left\{\left[\sqrt{2p_{0}p_{2}}(1-e^{\textbf{i}2\varphi}) +\underline{p_1e^{\textbf{i}\varphi}-p_1e^{\textbf{i}\varphi}}\right]\ket{00}+\sqrt{2p_{1}p_{2}}\left(1+\underline{e^{\textbf{i}\varphi}-e^{\textbf{i}\varphi}}\right)\ket{01}\right. \\
&\left. +\sqrt{2p_1p_2}\left(\underline{e^{\textbf{i}\varphi}-e^{\textbf{i}\varphi}} -e^{\textbf{i}2\varphi}\right)\ket{10} +2p_2\left( \underline{e^{\textbf{i}\varphi}-e^{\textbf{i}\varphi}}\right)\ket{11} %\\
%&\left.
+\sqrt{2}p_{2}\left(\ket{02}-e^{\textbf{i}2\varphi}\ket{20}\right)\right\},
\end{aligned}
\end{equation}
\end{widetext}
where $a'_{t-\tau}$ and $a'_{t}$ act in quantum states $\ket{\phi_{t-\tau}}$ and $\ket{\phi_{t}}$ respectively. 
If photons are all identical, all underlined terms cancel out, leaning to 
\begin{equation}
	\begin{aligned}
	\mathcal{C}(0)&=\frac{1}{4}p_{0}p_{2}(1-\cos2\varphi)+\frac{1}{4}p_1p_2+\frac{1}{4}p_2^2 \\
    &=\frac{p_{2}}{4}\left(1-p_{0}\cos2\varphi \right).
\end{aligned}
\end{equation}
%\end{widetext}
When considering finite photon indistinguishability defined in equation~\eqref{M:defination}, all terms that depend on $\varphi$ must be corrected accordingly and we then obtain 
\begin{equation}
	\begin{aligned}
	\mathcal{C}(0)=\frac{p_2}{4}\left(1-p_0 M \cos2\varphi \right)+\frac{p_1^2+4p_1p_2+4p_2^2}{8}\left( 1-M^\prime \right),
	\end{aligned}
  \label{eq:0_delay_coincidence}
\end{equation}
where $M^\prime$ has similar definition as $M$ but further takes into account for the detector time resolution. Effectively, we let $|\underline{e^{\textbf{i}\varphi}-e^{\textbf{i}\varphi}}|^2=2(1-M')$ in obtaining equation~(\ref{eq:0_delay_coincidence}).